\documentclass[12pt,letter]{article}

\usepackage{graphicx, epsfig, color}
\usepackage{amsmath,amssymb,hyperref}
\usepackage{subfigure}

\def\eslt{\not\!\!{E_T}}
\def\to{\rightarrow}

\def\bi{\begin{itemize}}
\def\ei{\end{itemize}}
\def\te{\tilde e}
\def\ta{\tilde a}

\def\tu{\tilde u}
\def\sps1ap{SPS1a$^\prime$}
\def\c1p{C1$^\prime$}

\def\ts{\tilde s}
\def\tb{\tilde b}
\def\tf{\tilde f}

\def\tst{\tilde t}
\def\ttau{\tilde \tau}

\def\tg{\tilde g}
\def\tnu{\tilde\nu}

\def\tw{\widetilde W}
\def\tz{\widetilde Z}
\def\alt{\lesssim}
\def\agt{\gtrsim}
\def\be{\begin{equation}}  
\def\ee{\end{equation}}  
\def\bea{\begin{eqnarray}}  
\def\eea{\end{eqnarray}}  
\def\beas{\begin{eqnarray*}}  
\def\eeas{\end{eqnarray*}}  
\def\delew{\Delta_{\rm EW}}
\newcommand\prd[3]{{\it Phys.\ Rev.\ }{\bf D #1} (#2) #3}

\newcommand\prl[3]{{\it Phys.\ Rev.\ Lett.\ }{\bf #1} (#2) #3}
\newcommand\plb[3]{{\it Phys.\ Lett.\ }{\bf B #1} (#2) #3}
\newcommand\jhep[3]{{\it J. High Energy Phys.\ }{\bf #1} (#2) #3}

\newcommand\npb[3]{{\it Nucl.\ Phys.\ }{\bf B #1} (#2) #3}

\newcommand{\hepph}[1]{hep-ph/#1}

\begin{document}
\begin{titlepage}
\begin{flushright}
OUHEP-170430,\ UH-511-1276-17
\end{flushright}
\vspace{0.3cm}
\begin{center}
{\Large \bf Superparticle phenomenology
 from the natural mini-landscape }\\
\vspace{1.2cm} \renewcommand{\thefootnote}{\fnsymbol{footnote}}
{\large Howard Baer$^1$\footnote[1]{Email: baer@nhn.ou.edu }, 
Vernon Barger$^2$\footnote[2]{Email: barger@pheno.wisc.edu },
Michael Savoy$^1$\footnote[3]{Email: savoy@ou.edu },\\
Hasan Serce$^1$\footnote[4]{Email: serce@ou.edu } and 
Xerxes Tata$^3$\footnote[5]{Email: tata@phys.hawaii.edu }
}\\ 
\vspace{1cm} \renewcommand{\thefootnote}{\arabic{footnote}}
{\it 
$^1$Department of Physics and Astronomy,
University of Oklahoma, Norman, OK 73019, USA \\
}
{\it 
$^2$Department of Physics,
University of Wisconsin, Madison, WI 53706, USA \\
}
{\it 
$^3$Department of Physics and Astronomy,
University of Hawaii, Honolulu, HI 96822, USA \\
}

\end{center}

\vspace{0.5cm}
\begin{abstract}
\noindent 

The methodology of the heterotic {\it mini-landscape} attempts to zero in on
phenomenologically viable corners of the string landscape where the
effective low energy theory is the Minimal Supersymmetric Standard Model
with localized grand unification. The gaugino mass pattern is that of
mirage-mediation. The magnitudes of various SM Yukawa couplings point to
a picture where scalar soft SUSY breaking terms are related to the
geography of fields in the compactified dimensions. Higgs fields and
third generation scalars extend to the bulk and occur in split
multiplets with TeV scale soft masses. First and second generation
scalars, localized at orbifold fixed points or tori with enhanced
symmetry, occur in complete GUT multiplets and have much larger
masses. This picture can be matched onto the parameter space of
generalized mirage mediation.  Naturalness considerations, the
requirement of the observed electroweak symmetry breaking pattern, and
LHC bounds on $m_{\tg}$ together limit the gravitino mass to the
$m_{3/2}\sim 5-60$ TeV range.  The mirage unification scale is bounded
from below with the limit depending on the ratio of squark to gravitino
masses. We show that while natural SUSY in this realization may escape
detection even at the high luminosity LHC, the high energy LHC with
$\sqrt{s}=33$~TeV could unequivocally confirm or exclude this
scenario. It should be possible to detect the expected light higgsinos
at the ILC if these are kinematically accessible, and possibly also
discriminate the expected compression of gaugino masses in the natural
mini-landscape picture from the mass pattern expected in models with
gaugino mass unification. The  thermal WIMP signal 
should be accessible via direct detection searches at 
the multi-ton noble liquid detectors such as XENONnT or LZ.

\vspace*{0.8cm}

\end{abstract}

\end{titlepage}

\section{Introduction}

String theory offers a UV complete finite theory which includes a
quantum mechanical treatment of gravity and the possible inclusion of
the Standard Model (SM)\cite{Schellekens:2013bpa}.  While only a few
string theories exist, formulated as 10-dimensional superstring or
11-dimensional $M$-theory, the compactification of the
extra-dimensions leads to a vast landscape for 4-D theories.  It
appears that neither the SM nor its minimal supersymmetric extension
are a {\em generic} part of the landscape. There has, nevertheless,
been a vast effort to understand how either of these models might
emerge from the landscape of string vacua\cite{Dienes:2015xua}.

One promising approach has been to adopt the SM as a low energy target
effective field theory and to see if it might arise in special regions
of the string landscape.  By investigating these so-called ``fertile
patches'' of the landscape, lessons may be learned about how the SM
might emerge from string theory compactification\cite{Nilles:2015wua}.
Since string theory necessarily involves a high mass scale 
$M_{\rm  string}$ close to $m_{\rm Pl}$ or $m_{\rm GUT}$, low energy ($N=1$)
supersymmetry (SUSY) is usually invoked to stabilize the Higgs mass
\cite{Witten:1981nf}, and
the low energy target effective
theory is frequently taken as the Minimal Supersymetric Standard Model
(MSSM).  The MSSM enjoys indirect phenomenological support in that
1. the measured values of weak scale gauge couplings unify under MSSM
renormalization group running at $Q=m_{\rm GUT}\simeq 2\times 10^{16}$
GeV\cite{gauge}, 2. the measured value of $m_t$ is just what is needed
to drive a radiative breakdown of electroweak symmetry\cite{rewsb}, and
3. the measured value of the Higgs boson mass $m_h\simeq 125$
GeV\cite{lhc_h} falls squarely within the required MSSM range where
$m_h\alt 135$ GeV is required\cite{mhiggs}.\footnote{
For some related approaches, see \cite{Cleaver:1998im,Cvetic:2002qa,
Kane:2006yi}.}

A very practical avenue for linking string theory to weak scale
physics, known as the mini-landscape, has been investigated at some
length \cite{Lebedev:2006kn}. 
The methodology of the mini-landscape is to zero in on 
the small subset of landscape vacua which give rise to reasonable
weak scale particle physics as realized by the MSSM.
The mini-landscape adopts the
$E_8\times E_8$ gauge structure of the heterotic string since one of
the $E_8$ groups automatically contains as sub-groups the grand unified
structures that the SM multiplets and quantum numbers seems to
reflect: $E_8\supset E_6\supset SO(10)\supset SU(5)\supset G_{\rm SM}$
where $G_{\rm SM}\equiv SU(3)_C\times SU(2)_L\times U(1)_Y$.  The other
$E_8$ may contain a hidden sector with $SU(n)$ subgroups which become
strongly interacting at some intermediate scale $\Lambda\sim 10^{13}$
GeV leading to gaugino condensation and consequent supergravity
breaking\cite{inocond}.  Compactification of the heterotic string on a
$Z_6-II$ orbifold\cite{Kobayashi:2004ya} can lead to low energy
theories which include the MSSM, possibly with additional exotic matter
states.

A detailed exploration of the mini-landscape has been performed a
number of years ago. In this picture, the properties of the 4-D low
energy theory are essentially determined by the geometry of the
compact manifold, and by the location of the matter superfields on
this manifold. The gauge group is $G_{\rm SM}$ though the symmetry may be
enhanced for fields confined to fixed points, or to fixed tori, in the
extra dimensions. Examination of the models which lead to MSSM-like
structures revealed the following picture\cite{Nilles:2015qka}.
\begin{enumerate}
\item The first two generations of matter live on orbifold fixed
  points which exhibit the larger $SO(10)$ gauge symmetry; thus, first
  and second generation fermions fill out the 16-dimensional spinor
  representation of $SO(10)$.
\item The Higgs multiplets $H_u$ and $H_d$ live in the untwisted
  sector and are bulk fields that feel just $G_{\rm SM}$.
  As such, they (and the gauge bosons) come in incomplete
  GUT multiplets which automatically solves the classic doublet-triplet
  splitting problem.
\item The third generation quark doublet and the top singlet also
  reside in the bulk, and thus have large overlap with the Higgs fields
  and correspondingly large Yukawa couplings. The location of other
  third generation matter fields is model dependent.  The small
  overlap of Higgs and first/second generation fields (which do not
  extend into the bulk) accounts for their much smaller Yukawa
  couplings.
\item Supergravity breaking may arise from hidden sector gaugino
  condensation with $m_{3/2}\sim \Lambda^3/m_{\rm Pl}^2$ with the gaugino
  condensation scale $\Lambda\sim 10^{13}$ GeV. SUSY breaking effects
  are felt differently by the various MSSM fields as these are located
  at different places in the compact manifold. Specifically, the Higgs
  and top squark fields in the untwisted sector feel extended
  supersymmetry (at tree level) in 4-dimensions, and are thus more
  protected than the fields on orbifold fixed points which receive protection
  from just $N=1$ supersymmetry \cite{Krippendorf:2012ir}. 
First/second
generation matter scalars are thus expected with masses $\sim m_{3/2}$. 
Third generation and Higgs soft mass parameters (which enjoy the
added protection from extended SUSY) are suppressed by an additional
loop factor $\sim 4\pi^2 \sim \log (m_{\rm Pl}/m_{3/2})$. 
Gaugino masses and trilinear soft terms are
expected to be suppressed by the same factor. 
The suppression of various soft SUSY breaking terms means that 
(anomaly-mediated) loop contributions\cite{amsb} may be comparable 
to modulus- (gravity-) mediated contributions leading to models with mixed
moduli-anomaly mediation\cite{choi} (usually dubbed as 
{\it mirage mediation} or MM for short);
in these scenarios, gaugino masses apparently unify at some
intermediate scale
\be
\mu_{\rm mir}\sim m_{\rm GUT}e^{-8\pi^2/\alpha}, 
\label{eq:mumir}
\ee
where $\alpha$ parametrizes the relative amounts of moduli- versus anomaly-mediation. 
\end{enumerate} 

The phenomenon of mirage mediation was originally found to occur in 
type-IIB strings where moduli fields were stabilized by a combination of 
fluxes and gaugino condensation leading to theories with an 
AdS vacuum. 
Uplifting of the AdS to a deSitter vacuum was arranged via the presence of
anti-$D3$ branes (KKLT formulation\cite{kklt}). 
Since these original models were first written down, additional uplifting 
mechanisms have been formulated\cite{uplift}. 
The mirage mediation SUSY breaking
scheme was also found to arise in compactification of the heterotic string
in addition to the original II-B proposal\cite{Lowen:2008fm}.

The mirage mediation soft SUSY breaking Lagrangian terms have been computed in 
a number of papers and are given by\cite{choi,choi3,flm},
\begin{eqnarray}
M_a&=& M_s\left(\alpha +b_a g_a^2\right),\label{eq:M}\\
A_{ijk}&=& M_s \left( -a_{ijk}\alpha +\gamma_i +\gamma_j +\gamma_k\right),
\label{eq:A}\\
m_i^2 &=& M_s^2\left( c_i\alpha^2 +4\alpha \xi_i -
\dot{\gamma}_i\right) ,\label{eq:m2}
\end{eqnarray}
where $M_s\equiv\frac{m_{3/2}}{16\pi^2}$,
$b_a$ are the gauge $\beta$ function coefficients for gauge group $a$ and 
$g_a$ are the corresponding gauge couplings. The coefficients that
appear in (\ref{eq:M})--(\ref{eq:m2}) are given by
$c_i =1-n_i$, $a_{ijk}=3-n_i-n_j-n_k$ and
$\xi_i=\sum_{j,k}a_{ijk}{y_{ijk}^2 \over 4} - \sum_a g_a^2
C_2^a(f_i).$ 
Finally, $y_{ijk}$ are the superpotential Yukawa couplings,
$C_2^a$ is the quadratic Casimir for the a$^{th}$ gauge group
corresponding to the representation to which the sfermion $\tf_i$ belongs,
$\gamma_i$ is the anomalous dimension, and
$\dot{\gamma}_i =8\pi^2\frac{\partial\gamma_i}{\partial \log\mu}$.
Expressions for the last two quantities involving the 
anomalous dimensions can be found in the Appendix of Ref's~\cite{flm,cjko}.

In the earliest models the coefficients that appear in
(\ref{eq:A}) and (\ref{eq:m2}) took on values determined by discrete
values of the modular weights $n_i$ which depended on the location of
fields in the original II-B string model and were given by $c_i
=1-n_i$, $a_{ijk}=3-n_i-n_j-n_k$.
Thus, the parameter space of the original MM models was
given by 
\be 
m_{3/2},\ \alpha ,\ \tan\beta ,\ sign(\mu ),\ n_i.
\label{eq:kkltparam}
\ee 

It has since been recognized that while the gaugino mass patterns in
Eq.~(\ref{eq:M}) are a robust prediction of the mirage-mediation
picture, the corresponding patterns of scalar mass and trilinear
parameters are sensitive to the mechanisms of moduli stabilization and
uplifting. This, together with the fact that the original
mirage-mediation models seemed to require relatively large fine-tuning
in light of the measured value of the Higgs boson mass\cite{Baer:2014ica}, 
led us to suggest a phenomenological generalization of this picture 
discussed in Sec.~\ref{sec:nGMM} \cite{Baer:2016hfa}. 
This extension allows us to accommodate the mass
patterns suggested by the mini-landscape picture mentioned above, the
phenomenology of which is the subject of this paper. In
Sec.~\ref{sec:natmini} we explore the parameter space of this
generalized mirage mediation (GMM) framework, identify portions
which are consistent with naturalness, and study the resulting sparticle spectra
expected in the natural mini-landscape picture.  In
Sec.~\ref{sec:scan}, we perform scans over parameter space to obtain
upper bounds on superpartner masses from naturalness
requirements. Sec.~\ref{sec:col_dm}, we broadly discuss collider and
dark matter phenomenology of the natural mini-landscape picture. We summarize
our main results in Sec.~\ref{sec:conclude}.

\section{Natural generalized mirage mediation}
\label{sec:nGMM}

We have just mentioned that the original MM models
based on the parameter space
(\ref{eq:kkltparam}) were found to be
highly  fine-tuned even under the most conservative electroweak fine-tuning
measure $\delew$\cite{Baer:2012up,Baer:2012cf} for parameter
choices which gave rise to $m_h\sim 123-127$ GeV\cite{Baer:2014ica}.
The electroweak fine-tuning measure $\delew$ is defined by
requiring that there are no large cancellations between independent
 contributions to the $Z$ boson mass calculated from the minimization
 conditions of the 1-loop MSSM scalar potential,
%
\bea
\frac{m_Z^2}{2}&=&\frac{m_{H_d}^2+\Sigma_d^d-(m_{H_u}^2+\Sigma_u^u)\tan^2\beta}{\tan^2\beta -1}-\mu^2. \label{eq:mzs}
\eea
Here $\Sigma_u^u$ and $\Sigma_d^d$ denote 1-loop corrections
(expressions can be found in the Appendix of Ref. \cite{Baer:2012cf}) to the
scalar potential, $m_{H_u}^2$ and $m_{H_d}^2$ are the weak scale 
values of the soft breaking Higgs masses
and $\tan \beta \equiv \langle H_u \rangle / \langle H_d \rangle$.
SUSY models requiring large cancellations between the various terms on the
right-hand-side of Eq.~(\ref{eq:mzs}) to reproduce the measured value of
$m_Z^2$ are regarded as unnatural, or fine-tuned.
Thus, natural SUSY models are characterized by low values of the {\it
  electroweak} naturalness measure $\delew$ defined as
\cite{Baer:2012up,Baer:2012cf} 
\be \delew\equiv \text{max}|{\rm
  each\ term\ on\ RHS\ of\ Eq.}~(\ref{eq:mzs})|/(m_Z^2/2).  \ee

It is essential that the sensitivity of $m_Z$ be evaluated only with
respect to the {\it independent} parameters of the theory. If this is
not done, the UV sensitivity of the theory will be over-estimated, and
the theory may be incorrectly regarded as fine-tuned.  It has further
been shown that traditionally used high scale measures of
fine-tuning\cite{Barbieri:1987fn,Kitano:2006gv,Papucci:2011wy} reduce
to $\delew$ once underlying (potential) correlations between
parameters are properly
incorporated\cite{Baer:2013gva,Mustafayev:2014lqa,Baer:2014ica}. For
this reason, we regard $\delew$ as the most conservative
measure of fine-tuning.

It seems highly implausible that if the SUSY breaking parameter
$m_{H_u}^2$ runs large negative such that $-m_{H_u}^2\gg m_Z^2$, then
the value of the SUSY-conserving parameter $\mu$, which likely has a
very different origin from the soft terms, would be of just the
right value to nearly cancel against $-m_{H_u}^2$ and yield the (much
smaller) observed value of $m_Z$.
{\it Electroweak naturalness} then implies that
\bi
\item $m_{H_u}^2\sim -(100-300)^2$ GeV$^2$, and
\item $\mu^2\sim (100-300)^2$ GeV$^2$\cite{Chan:1997bi,Baer:2011ec}
  \ei
(the closer to $m_Z$ the better).
For moderate-to-large $\tan\beta \gtrsim 5$, the remaining contributions
other than $\Sigma_u^u$ are suppressed.  The largest radiative
corrections $\Sigma_u^u$ typically come from the top squark
sector. The value of the trilinear soft term $A_0 \sim -1.6m_0$ leads to
TeV-scale top squarks and minimizes $\Sigma_u^u(\tst_{1,2})$ while
simultaneously lifting the Higgs mass $m_h$ to $\sim 125$~GeV\cite{Baer:2012cf}. 

The failure of naturalness in MM as detailed above has led us
previously to propose moving from discrete choices of the parameters
$a_{ijk}$ and $c_i$ in Eqs.~(\ref{eq:A}) and~(\ref{eq:m2}) to a
continuous range, and also to allow $c_i$ values greater than
1\cite{Baer:2016hfa}.  While the discrete parameter choices occur in a
wide range of KKLT-type compactifications (for some discussion, see
Ref.~\cite{choi_sax}), a continuous range of these parameters may be
expected if one allows for more general methods of moduli
stabilization and potential uplifting.  For instance, if the soft
terms scan as in the string landscape picture\cite{Douglas:2012bu},
then their moduli-mediated contributions may be expected to be
parametrized by a continuous value.  For models which generate a small
$\mu$ term $\sim 100$ GeV from multi-TeV soft terms, such as in the
Kim-Nilles mechanism\cite{Kim:1983dt} with radiative Peccei-Quinn
breaking\cite{radpq}, it has been suggested that the statistical pull
by the landscape towards large soft terms, coupled with the anthropic
requirement of $m_{\rm weak}\sim 100$ GeV, acts as an attractor
towards natural SUSY soft term boundary conditions\cite{landscape}.

Note that the phenomenological modification we have suggested will not
affect the result Eq.~(\ref{eq:M}) for gaugino mass parameters, which has
been stressed\cite{Choi:2007ka} to be the most robust prediction of
the MM mechanism.  In this paper, we allow for the more {\it general}
mirage mediation (GMM) parameters, thus adopting a parameter space
given by 
\be 
\alpha,\ m_{3/2},\ c_m,\ c_{m3},\ a_3,\ c_{H_u},\ c_{H_d},\
\tan\beta \ \  \ \ (GMM), 
\ee 
where $a_3$ is short for $a_{Q_3H_uU_3}$.
Here, we adopt an independent value $c_m$ for the first two
matter-scalar generations whilst the parameter $c_{m3}$ applies to
third generation matter scalars. This splitting accomodates the case
of the mini-landscape wherein third generation scalars are expected to
receive soft terms $\sim$ TeV whilst first/second generation matter
scalars are expected to occur with mass values $\sim m_{3/2}\gg 1$
TeV.  The independent values of $c_{H_u}$ and $c_{H_d}$ which set the
moduli-mediated contribution to the soft Higgs mass terms may
conveniently be traded for weak scale values of $\mu$ and $m_A$ as is
done in the two-parameter non-universal Higgs model (NUHM2)\cite{nuhm2}:
\be
\alpha,\ m_{3/2},\ c_m,\ c_{m3},\ a_3,\ \tan\beta , \mu ,\ m_A
\ \ \ (GMM^\prime ). \label{eq:gmmp}  \ee
This procedure allows for more direct
exploration of natural SUSY parameter space which requires $\mu\sim
100-300$ GeV (the closer to $m_Z$ the better).  Thus, our final
relevant soft terms are given by
\begin{eqnarray}
M_a&=& \left( \alpha +b_a g_a^2\right)m_{3/2}/16\pi^2,\label{eq:Ma}\\
A_{\tau}&=& \left( -a_3\alpha +\gamma_{L_3} +\gamma_{H_d} +\gamma_{E_3}\right)m_{3/2}/16\pi^2,\\
A_{b}&=& \left( -a_3\alpha +\gamma_{Q_3} +\gamma_{H_d} +\gamma_{D_3}\right)m_{3/2}/16\pi^2,\\
A_{t}&=& \left( -a_3\alpha +\gamma_{Q_3} +\gamma_{H_u} +\gamma_{U_3}\right)m_{3/2}/16\pi^2,\\
m_i^2(1,2) &=& \left( c_m\alpha^2 +4\alpha \xi_i -\dot{\gamma}_i\right)
(m_{3/2}/16\pi^2)^2 ,\label{eq:mi2} \\
m_j^2(3) &=& \left( c_{m3}\alpha^2 +4\alpha \xi_j -\dot{\gamma}_j\right)
(m_{3/2}/16\pi^2)^2 ,\\
m_{H_u}^2 &=& \left( c_{H_u}\alpha^2 +4\alpha \xi_{H_u} -\dot{\gamma}_{H_u}\right)
(m_{3/2}/16\pi^2)^2 ,\\
m_{H_d}^2 &=& \left( c_{H_d}\alpha^2 +4\alpha \xi_{H_d} -\dot{\gamma}_{H_d}\right)
(m_{3/2}/16\pi^2)^2 ,\label{eq:MHd}
\end{eqnarray}
where, for a given value of $\alpha$ and $m_{3/2}$, the values of
$c_{H_u}$ and $c_{H_d}$ are adjusted so as to fulfill the input values
of $\mu$ and $m_A$. In the above expressions, the index $i$ runs over
first/second generation MSSM scalars
$i=Q_{1,2},U_{1,2},D_{1,2},L_{1,2}$ and $E_{1,2}$ while $j$ runs overs
third generation scalars $j=Q_3,U_3,D_3,L_3$ and $E_3$. 
The common value of $c_m$ in Eq.~(\ref{eq:mi2}) ensures that
flavor-changing neutral current (FCNC) processes are suppressed.
The GMM parameter space is well-suited for
the exploration of the superparticle mass spectra and resulting
phenomenology that is to be expected from the natural
mini-landscape. With this in mind, we have recently included the GMM
model as model line \#12 into the event generator program Isajet
7.86\cite{isajet}.

\section{Superparticle spectra from the natural mini-landscape}
\label{sec:natmini}

\begin{figure}[tbp]
\begin{center}
\includegraphics[height=0.4\textheight]{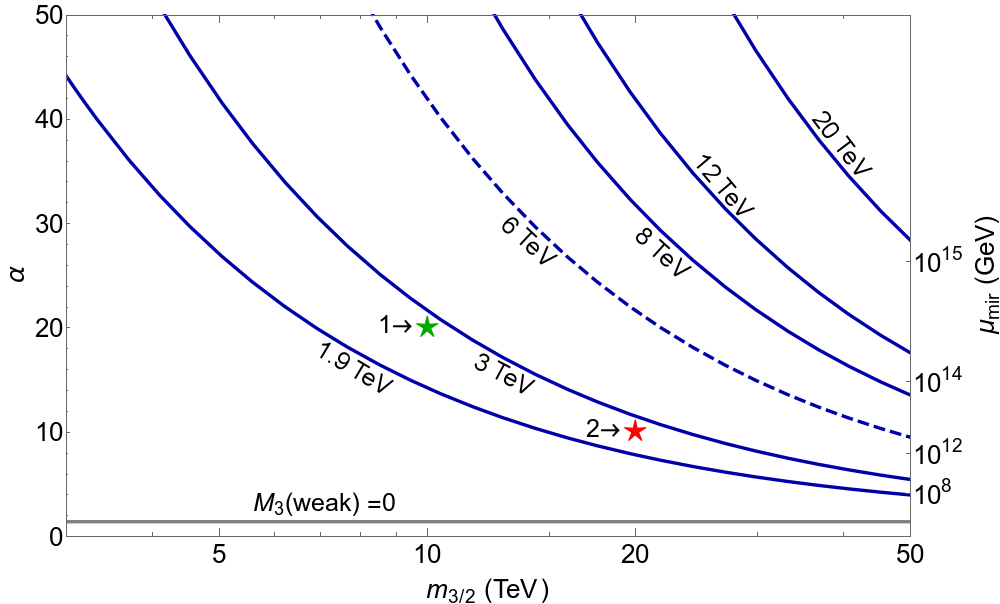}
\caption{Contours of $M_3({\rm weak})$ in the $\alpha$ vs. $m_{3/2}$ plane of
  the GMM model.
  The region below $M_3\sim 1.9$ is excluded by LHC gluino
  pair searches. The locations of the benchmark points mini1 and mini2
  are shown by green and red stars, respectively.  The region
  below the dashed $M_3\sim 6$ TeV contour has the capacity to be
  natural.  On the right side, some corresponding values of $\mu_{mir}$
  are shown.
\label{fig:M3}
}
\end{center}
\end{figure}

We begin by reminding the reader that in the natural mini-landscape
picture, 1.~the gaugino mass spectrum is as given by mirage
mediation Eq.~(\ref{eq:Ma}), 2.~$|\mu|$ not far from $m_Z$, 
3.~third generation squarks lie in the TeV range, and
4.~first and second generation masses are close to $m_{3/2}\sim$ multi-TeV.
To obtain a broad overview, we show
in Fig. \ref{fig:M3} the value of $M_3({\rm weak})$ (where $m_{\tg}\sim
M_3({\rm weak})$ up to loop corrections) as generated using
Eq.~(\ref{eq:Ma}) -- but scaled by a factor $M_3({\rm weak})\simeq 2.34
M_3({\rm GUT})$ to account roughly for RG evolution -- without making a
specific assumption about scalar sector parameters. 
From the figure, we immediately see that the LHC13 limit 
$m_{\tg} \alt 1.9$~TeV\cite{atlas_s,cms_s},
roughly speaking, excludes values of $\alpha$ below the
$M_3({\rm weak})=1.9$ TeV contour.  Moreover, the fact that the naturalness
condition bounds the gluino mass from above similarly excludes values of
$\alpha$ above the dashed curve if one insists on EW naturalness with
$\delew<30$ \cite{Baer:2017yqq}. We regard the large range of
$m_{3/2}$ and $\alpha$ between these curves as the ``favoured region''
of the mini-landscape picture, but keep in mind that the boundaries
have some fuzziness in part because the curves are only approximate
contours of the gluino mass. We will see later that -- for natural
sparticle mass spectra from the mini-landscape -- $m_{3/2}$ is in
fact bounded from above, the exact bound depending on the details of the
mini-landscape picture.
%

\subsection{A natural mini-landscape benchmark point}

To gain some perspective on natural mini-landscape parameter space, we
first generate a benchmark (BM) point using Isajet 7.86. We adopt a
value $m_{3/2}=10$ TeV and then select a value of $\alpha=20$ well
within the allowed region of Fig. \ref{fig:M3}, the location of which
is shown by the green star.  To obtain the first two generation mass
parameters $\sim m_{3/2}$ we choose $c_m=100$ in Eq.~(\ref{eq:mi2}).
This leads to first/second generation soft terms $\sim 12.7$~TeV.  To
gain third generation scalars in the several TeV range, we select
$c_{m3}=18$ leading to $m_i(3)\sim 5.4$ TeV.  A choice of $a_3\sim 6$
leads to a GUT scale trilinear soft term $A_t\sim -7.6$ TeV which is a
typical value required to boost the Higgs mass $m_h$ up to its
measured value $\sim 125$ GeV\cite{h125} whilst simultaneously
reducing $\delew$ to natural values\cite{Baer:2012up}. In addition, we
choose a natural value of $\mu =150$ GeV, with $\tan\beta =10$ and
$m_A=2$ TeV.  The sparticle spectrum from Isajet 7.86 is listed in
Table \ref{tab:bm} as the  BM point mini1. The spectrum for
an NUHM3 model that should be in close correspondence with the BM
mini1 point is shown in the adjacent column of this
table.\footnote{The NUHM3 model is a three parameter extension of the
  familiar mSUGRA/CMSSM model in which the two GUT scale Higgs mass
  parameters as well as the GUT scale third generation sfermion mass
  parameters are taken to be independent of the universal scalar mass
  $m_0$ of the mSUGRA framework. The mini-landscape picture is then
  very close to the NUHM3 picture except that the GUT scale gaugino
  mass pattern is given by mirage mediation, and that scalar masses
  and $A$-parameters include small anomaly-mediated contributions.}
The last column lists the spectrum and parameters of a second
mini-landscape point introduced in Sec.~\ref{subsubsec:mini2}. From
the Table, we see that for the mini1 BM point the first/second
generation matter scalars lie at $m_i(1,2)\sim 12.8$ TeV while third
generation scalars are in the several TeV range with $m_{\tst_1}=1564$
GeV.  The gluino comes in at $2.9$ TeV. Both $m_{\tg}$ and
$m_{\tst_1}$ are well above current LHC13 limits. The Higgs mass at
124.3 GeV is in accord with its measured value if one allows for a
$\pm 2$ GeV theory error in the Isajet computation of
$m_h$. The value of $\delew=11.8$ or
$\delew^{-1}=8.5\%$ fine-tuning.  Thus, the spectrum of the mini1
benchmark model is very natural and the underlying string model
that results in the mini-landscape picture with 
the chosen values of $c_{m3}$ and $a_3$ would yield a natural
high scale theory.  The thermally-produced (TP) relic density of
higgsino-like WIMPs comes in at $\Omega_{\tz_1}^{TP}h^2=0.007$, below
the measured value by a factor 17.  The remainder may be made up by
other particles:
since we also insist on naturalness in the QCD sector, the QCD axion
is the likely candidate.  The  relic abundance of both the QCD
axion and higgsino-like WIMPs depends on various parameters from the
Peccei-Quinn sector (axino and saxion mass, axion mis-alignment angle,
axion decay constant $f_a$ etc.)\cite{Bae:2013bva}.
%
\begin{table}\centering
\begin{tabular}{lccc}
\hline
parameter & mini1 & NUHM3 & mini2\\
\hline
$m_{3/2}$      & 10000 & -- & 20000 \\
$\alpha$       & 20  & -- & 10 \\
$c_m$          & 100 & -- & 250  \\
$c_{m3}$          & 18 & -- & 23 \\
$a_3$           & 6 & -- & 6\\
$\tan\beta$    & 10  & 10 & 10 \\
\hline
$\mu$          & 150  & 150 & 150 \\
$m_A$          & 2000 & 2000 & 2000 \\
\hline
$m_{\tg}$   & 2911.5 & 2916.2 & 2784.5 \\
$m_{\tu_L}$ & 12810.5 & 12754.5 & 20097.5 \\
$m_{\tu_R}$ & 12888.2 & 12830.6 & 20177.8 \\
$m_{\te_R}$ & 12589.0 & 12525.1 & 19965.9  \\
$m_{\tst_1}$& 1564.5 & 1787.2 & 1341.7 \\
$m_{\tst_2}$& 3805.3 & 3869.5 & 3671.2  \\
$m_{\tb_1}$ & 3840.5 & 3899.5 & 3709.6 \\
$m_{\tb_2}$ & 5306.0 & 5321.7 & 5432.4 \\
$m_{\ttau_1}$ & 5097.3 & 5090.3 & 5757.6 \\
$m_{\ttau_2}$ & 5399.8 & 5386.1 & 5970.8 \\
$m_{\tnu_{\tau}}$ & 5373.3 & 5358.9 & 5933.1 \\
$m_{\tw_2}$ & 1132.3 & 1026.4 & 1178.5 \\
$m_{\tw_1}$ & 157.7 & 157.5 & 157.6 \\
$m_{\tz_4}$ & 1144.4 & 1038.5 & 1187.5 \\ 
$m_{\tz_3}$ & 674.0 & 537.6 & 773.7 \\ 
$m_{\tz_2}$ & 156.8 & 157.3 & 156.5 \\ 
$m_{\tz_1}$ & 148.5 & 147.3 & 148.8 \\ 
$m_h$       & 124.3 & 124.2 & 124.3 \\ 
\hline
$\Omega_{\tz_1}^{std}h^2$ & 0.007 & 0.007 & 0.006 \\
$BF(b\to s\gamma)\times 10^4$ & $3.1$ & $3.1$ & 3.1 \\
$BF(B_s\to \mu^+\mu^-)\times 10^9$ & $3.8$ & 3.8 & 3.8 \\
$\sigma^{SI}(\tz_1, p)$ (pb) & $1.1\times 10^{-9}$ & $1.5\times 10^{-9}$ &
$9.1\times 10^{-10}$  \\
$\sigma^{SD}(\tz_1 p)$ (pb) & $3.6\times 10^{-5}$ & $5.6\times 10^{-5}$ 
& $3.2\times 10^{-5}$  \\
$\langle\sigma v\rangle |_{v\to 0}$  (cm$^3$/sec)  & $3.0\times 10^{-25}$
& $3.0\times 10^{-25}$ & $3.0\times 10^{-25}$ \\
$\delew$ & 11.8 & 26.2 & 17.6 \\
\hline
\end{tabular}
\caption{Input parameters and masses in~GeV units for a natural
  mini-landscape SUSY benchmark point as compared to a similar point
  with gaugino mass unification from the NUHM3 model.  For the NUHM3 case we
  take $m_0(1,2)=12.6$ TeV, $m_0(3)=5360$ GeV, $m_{1/2}=1176$ GeV,
  $A_0=-7452$ GeV.  Also shown is the spectrum for a second
  mini-landscape point with $m_{3/2}=20$ TeV and $\alpha=10$.  We take
  $m_t=173.2$ GeV.  }
\label{tab:bm}
\end{table}
\begin{figure}[tbp]
\begin{center}
\includegraphics[height=0.23\textheight]{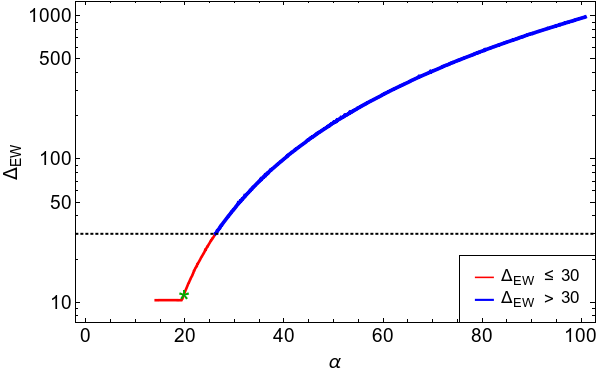}
\includegraphics[height=0.23\textheight]{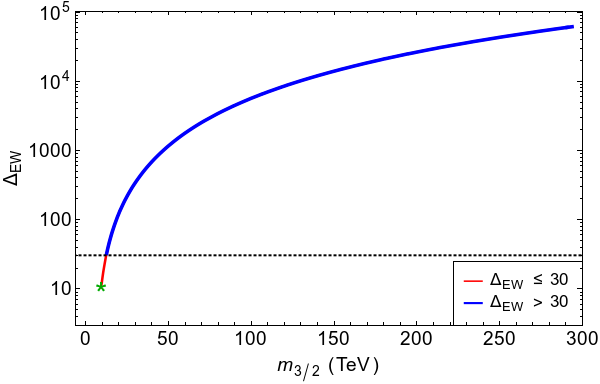}\\
\includegraphics[height=0.23\textheight]{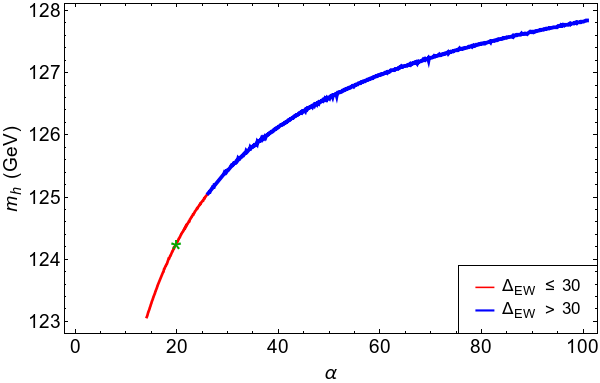}
\includegraphics[height=0.23\textheight]{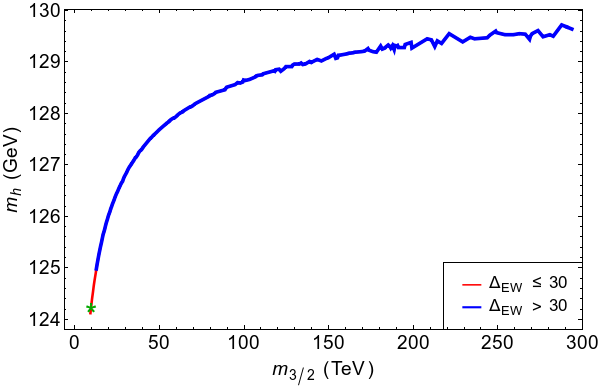}\\
\includegraphics[height=0.23\textheight]{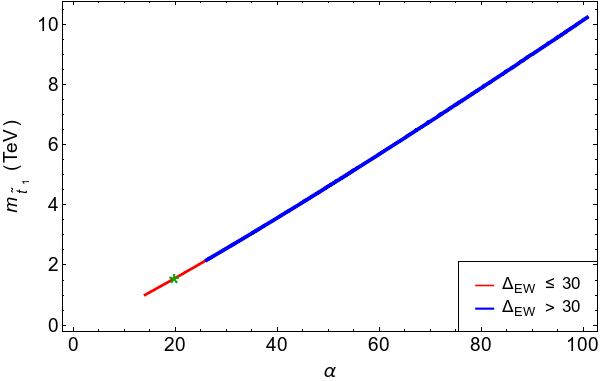}
\includegraphics[height=0.23\textheight]{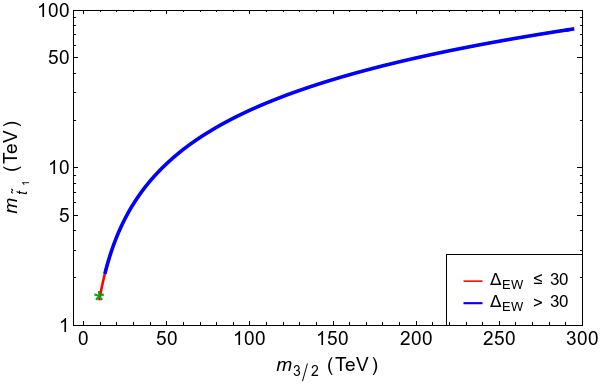}\\
\caption{$\delew$, $m_h$ and $m_{\tst_1}$  vs. variation in 
$\alpha$ and $m_{3/2}$ for the mini1 benchmark point.
The red portion of the curves has $\delew<30$.
The green star denotes the mini1 benchmark point.
\label{fig:al_m32}}
\end{center}
\end{figure}

To see how aspects of the mini1 benchmark point depend on $\alpha$ and
$m_{3/2}$, we show in Fig. \ref{fig:al_m32} the variation in
$\delew$, $m_h$ and $m_{\tst_1}$ versus $\alpha$ (left-column)
and versus $m_{3/2}$ (right-column) where other parameters remain
fixed at their mini1 BM values. The corresponding gluino mass can be
inferred from Fig.~\ref{fig:M3} while the higgsino masses are $\sim
|\mu|$. Other sfermions are typically too heavy to be produced at
LHC14. The red portion of the curves has $\delew<30$ and the
mini1 BM point is marked by the green cross.  In the upper left frame,
we see that $\delew$ rises rapidly with increasing $\alpha$ since
the superpartners (most notably the stops and gluino) become too heavy
and the spectrum becomes unnatural, even with $\mu$ fixed at $150$
GeV. This is due to the increasing values of radiative corrections,
mainly $\Sigma_u^u(\tst_{1,2}) $ in Eq.~(\ref{eq:mzs}).  Also, $m_h$ and
$m_{\tst_1}$ rise with increasing $\alpha$ as the top squarks become
increasingly heavy. Likewise, in the right column, we see
$\delew$ rapidly increases with increasing $m_{3/2}$, as do
$m_h$ and $m_{\tst_1}$. This is again due to rapidly increasing
sparticle masses.

\begin{figure}[tbp]
\begin{center}
\includegraphics[height=0.23\textheight]{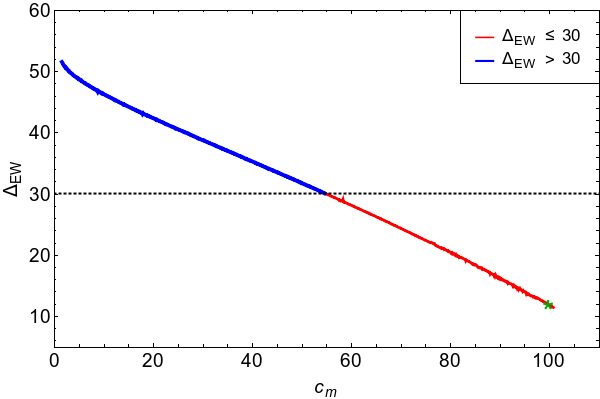}
\includegraphics[height=0.23\textheight]{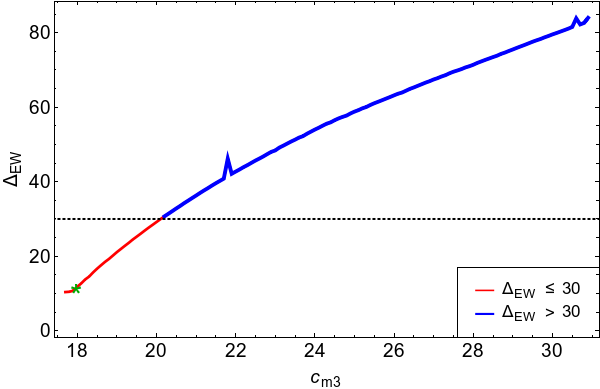}\\
\includegraphics[height=0.23\textheight]{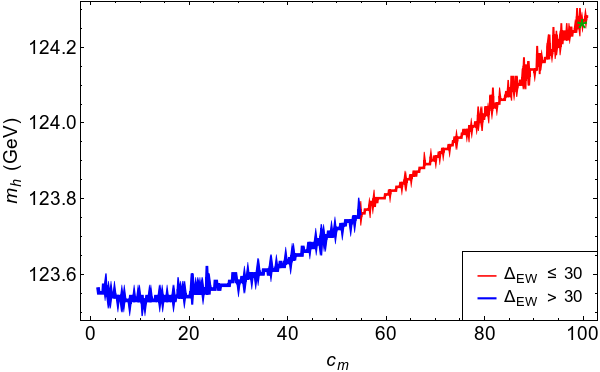}
\includegraphics[height=0.23\textheight]{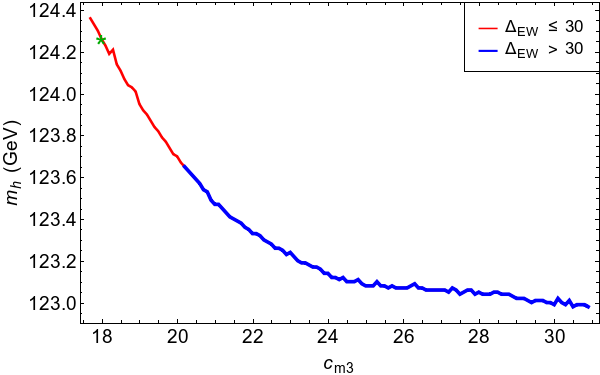}\\
\includegraphics[height=0.23\textheight]{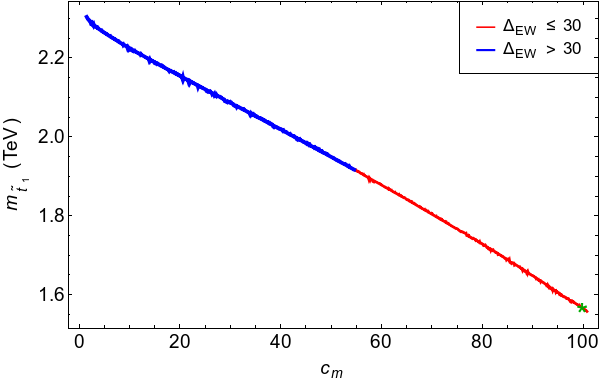}
\includegraphics[height=0.23\textheight]{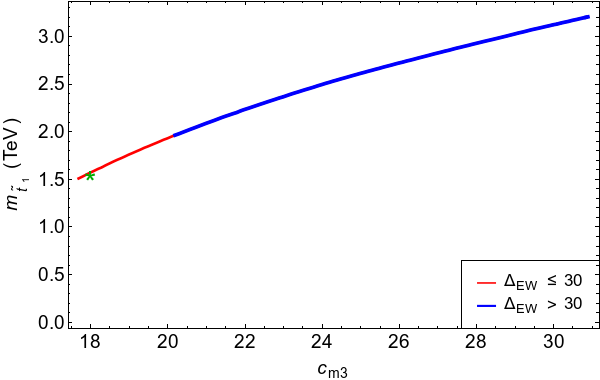}\\
\caption{$\delew$, $m_h$ and $m_{\tst_1}$  vs. variation in 
$c_m$ and $c_{m3}$ for the mini1 benchmark point.
The red portion of the curves has $\delew<30$.
The green star denotes the mini1 benchmark point.
\label{fig:cm_cm3}}
\end{center}
\end{figure}

In Fig. \ref{fig:cm_cm3}, we show variation in $\delew$, $m_h$ and
$m_{\tst_1}$ versus $c_m$ (left-column) and $c_{m3}$ (right-column),
this time holding $\alpha$ and $m_{3/2}$ fixed at their mini1
benchmark values.  From the upper-left frame, we see that $\delew$
rapidly {\it drops} with increasing $c_m$. At first thought, one might
not expect such sensitivity since $c_m$ governs first/second
generation scalar masses which seemingly have little to do with
naturalness. However, long ago it has been emphasized that two-loop RG
contributions\cite{mv} to scalar running become large for large
first/second generation scalar soft terms (see \cite{Baer:2000xa} and
more recently discussion in \cite{Krippendorf:2012ir}). These two loop
RGE terms help drive the stop sector towards natural values as seen in
the figure.  As elaborated later, this same RG evolution also leads to
a bound on the mini-landscape parameter space since too large values
for first/second generation scalars drive third generation soft mass
parameters tachyonic, leading to charge and color breaking (CCB)
minima in the scalar potential. For the mini1 BM point, viable spectra
are only generated out to $c_m \sim 100$, comfortably containing the
$\delew \le 30$ region.
In the right-column of Fig.~\ref{fig:cm_cm3}, we show how the same
quantities vary versus $c_{m3}$.  As $c_{m3}$ drops to values below
$\sim 17.5$, some top squark soft mass parameters are driven tachyonic
leading to CCB minima.
Larger values of $c_{m3}$ than are shown can be phenomenologically allowed, 
but only at an increasing cost to naturalness.

\begin{figure}[tbp]
\begin{center}
\includegraphics[height=0.4\textheight]{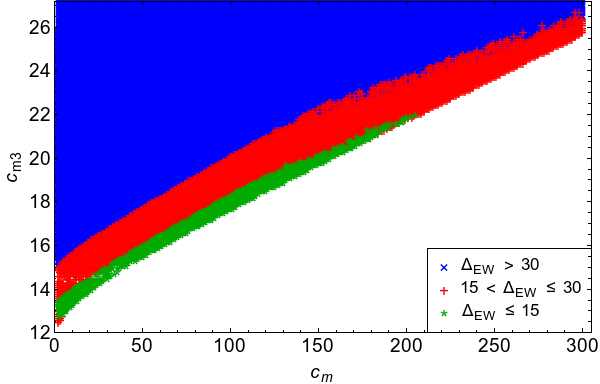}
\caption{Allowed (colored) points in $c_{m3}$ vs. $c_m$ plane.
\label{fig:cm3cm}}
\end{center}
\end{figure}

The interplay between the first/second and third generation scalar
masses is illustrated in the $c_{m3}$ vs. $c_m$ plane shown in
Fig. \ref{fig:cm3cm}, with other parameters fixed to their mini1 BM
values. We see again that as $c_m$ increases (for fixed $c_{m3}$), 
the model becomes increasingly {\it natural} as exhibited by lower values of
$\delew$ dropping below 15. For yet higher $c_m$ values,
solutions are rejected due to CCB minima mentioned above. Also, as
$c_{m3}$ drops, the solutions become increasingly natural. The
dividing line between natural (green) and forbidden (unshaded)
solutions corresponds to {\it barely-broken} electroweak symmetry which is the
essence of SUSY EW naturalness.  In References. \cite{Giudice:2006sn} and
\cite{landscape}, it is noted that  large $c_m$ solutions 
may be favoured by
a string theory landscape which prefers large soft terms, consistent with the
anthropic weak scale requirement $m_{W,Z,h}\sim 100$ GeV.

\begin{figure}[tbp]
\begin{center}
\includegraphics[height=0.23\textheight]{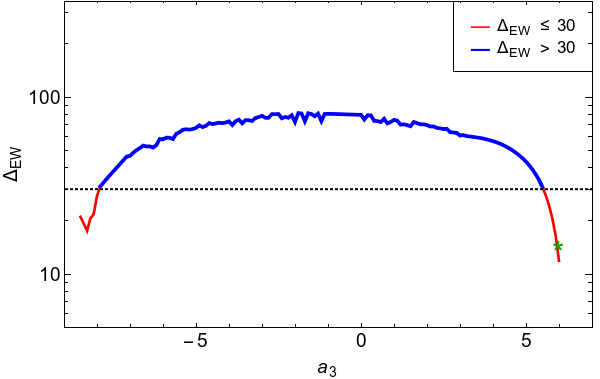}
\includegraphics[height=0.23\textheight]{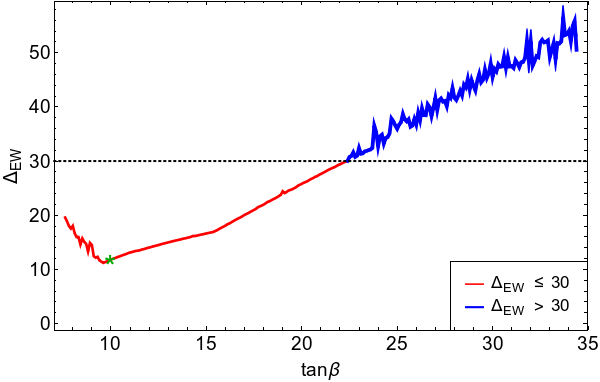}\\
\includegraphics[height=0.23\textheight]{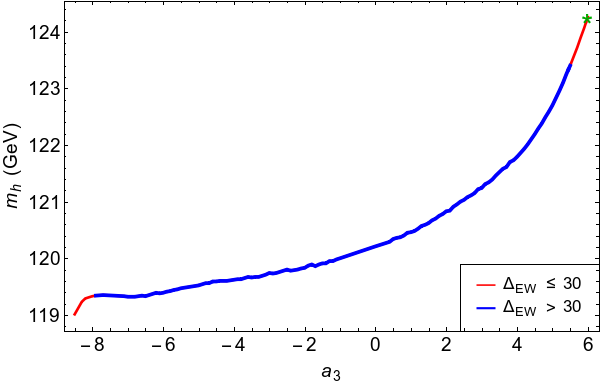}
\includegraphics[height=0.23\textheight]{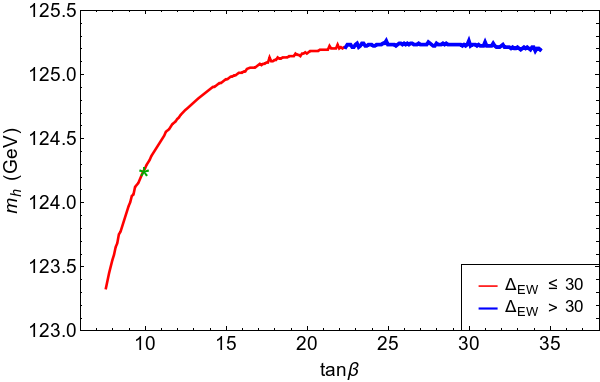}\\
\includegraphics[height=0.23\textheight]{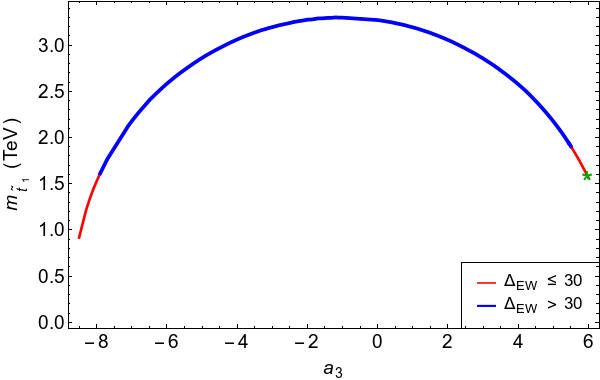}
\includegraphics[height=0.23\textheight]{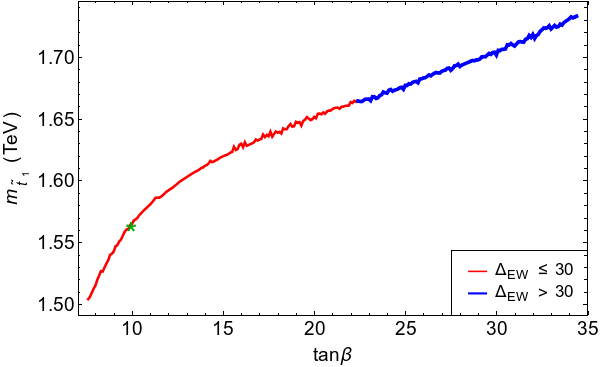}\\
\caption{$\delew$, $m_h$ and $m_{\tst_1}$  vs. variation in 
$a_3$ and $\tan\beta$ for the mini1 benchmark point.
The red portion of the curves has $\delew<30$.
The green star denotes the mini1 benchmark point.
\label{fig:a3_tanb}}
\end{center}
\end{figure}

In Fig. \ref{fig:a3_tanb}, we show variation of $\delew$, $m_h$ and
$m_{\tst_1}$ versus $a_3$ (left-column) and $\tan\beta$
(right-column).  For much of the range of $a_3$, which dictates the
magnitude of the trilinear soft terms $A_{t,b,\tau }$, the solutions
are relatively unnatural and the value of $m_h$ is too low.  For large
$a_3\sim 5-6$, both the mixing in the stop sector and radiative
corrections to $m_h$ increase, leading to $m_h\sim 125$ GeV whilst
simultaneously reducing $\delew<30$. The value of $m_h$ decreases for
negative values of $a_3$ because $A_t({\rm weak}) \sim 1$~TeV for $a_3
< -6$ to be compared with $A_t({\rm weak}) \sim -4$~TeV at the right
end of the plot.  The value of $m_{\tst_1}$ gets reduced for values of
$a_3$ consistent with both naturalness as well as the observed value of
$m_h$.  For large negative $a_3$, the value of $\delew$ also drops
below 30, but in this case $m_h$ remains around 119 GeV. From the
right-column, we see that low $\delew$ prefers low $\tan\beta
\alt25$. For higher $\tan\beta$, then the $b$-quark Yukawa increases
and the $\Sigma_u^u(\tb_{1,2})$ terms can contribute large values to
$\delew$ because the bottom squarks are typically heavier than the
stops.

\section{Scan over natural mini-landscape parameter space}\label{sec:scan}

In this section, we present results from a scan over the portion of
natural GMM parameter space which is in accord with expectations from
the mini-landscape. To facilitate the scanning, we first restrict the
high scale soft scalar mass parameters for the first two generations
(recall these have no protection from extended supersymmetry in 4D)
to be very close to $m_{3/2}$. Assuming that modulus-mediated
contributions dominate the soft terms,
we expect,
\be c_m\simeq
(16\pi^2/\alpha)^2 .
\label{eq:m1eqgrav}
\ee
(In Sec.~{\ref{subsec:extscan} below we will examine how our results vary if 
we relax this assumption.) 
Further, we will define $m_0(1,2)$ and $m_0(3)$ as the average 
of first/second and third generation matter scalars at the GUT scale.

\subsection{Results for $m_0(1,2)\simeq m_{3/2}$} \label{subsec:scan}
As mentioned, to begin our analysis we first present results taking
first/second generation SUSY breaking mass parameters close to the
gravitino mass, and scan over \bi
\item $\alpha :2-40$, 
\item $m_{3/2}: 3-65$ TeV,
\item $c_m:$ fixed at $(16\pi^2/\alpha)^2$ so that $m_0(1,2)\simeq m_{3/2}$,
\item $c_{m3}:1-{\rm min}\left[ 40,\ (c_m/4)\right]$
\item $a_3: 1-12$ in order to lift $m_h\sim 125$ GeV,
\item $\tan\beta :3-60$,
\item $\mu :100-360$ GeV (lower bound to enforce LEP2 chargino search limits
while upper limit from naturalness requiring $\delew<30$),
\item $m_A: 0.3-10$ TeV.
\ei
In addition, we require of our solutions
\bi
\item that there is an appropriate breakdown of EW symmetry 
({\it i.e.} EW breaks but with no CCB minima),
\item $m_h: 123-127$ GeV (allowing for $\sim \pm 2$ GeV theory error 
in our calculation of $m_h$),
\item $m_{\tg}>1.9$ TeV (in accord with recent LHC13 $\tg\tg$ search results),
\item $m_{\tst_1}>1$ TeV (in accord with recent LHC13
  $\tst_1\bar{\tst}_1$ search results\cite{atlas_t1,cms_t1}).  
\ei

The results of our scan are shown in Fig. \ref{fig:scan} where red
points have $\delew<30$ while green points have $\delew<20$.
From the plot we find an {\it upper bound} on $m_{3/2}\alt 24\ (32)$
TeV and $\delew < 20\ (30)$. For higher $m_{3/2}$ values,
first/second generation scalars are so heavy that some third
generation scalars always are driven tachyonic leading to CCB minima.
Since in the mini-landscape we expect $m_i(1,2)\sim m_{3/2}\sim \log
(m_{\rm Pl}/m_{3/2}) \times m_j(3)$ then we really expect $m_{3/2}\agt 5-10$ TeV.
The upper bound restricts the gravitino mass $m_{3/2}\alt 30$ TeV.
This has three effects on phenomenology: 1. We expect first/second
generation matter scalars to decouple from LHC searches, 2. the rather
high first/second generation scalars suppress possible FCNC and
CP-violating processes (offering at least a partial solution to the
SUSY flavor and CP problems)\cite{Dine:1990jd}, and 3. it softens the
cosmological gravitino problem wherein thermal production of
gravitinos followed by delayed decays can disrupt the successful
predictions of Big Bang nucleosynthesis (in that heavier gravitinos
decay more quickly and may then decay before the onset of
BBN)\cite{Khlopov:1984pf,Kohri:2005wn}.  Note that in these models the
moduli masses are expected to be $\sim \log (m_{\rm Pl}/m_{3/2}) m_{3/2}$ so
that for $m_{3/2}\sim 10-20$ TeV, then $m_T\sim 400-800$ TeV. Such
heavy moduli decay relatively rapidly and thus evade the cosmological
moduli problem.
%
\begin{figure}[tbp]
\begin{center}
\includegraphics[height=0.39\textheight]{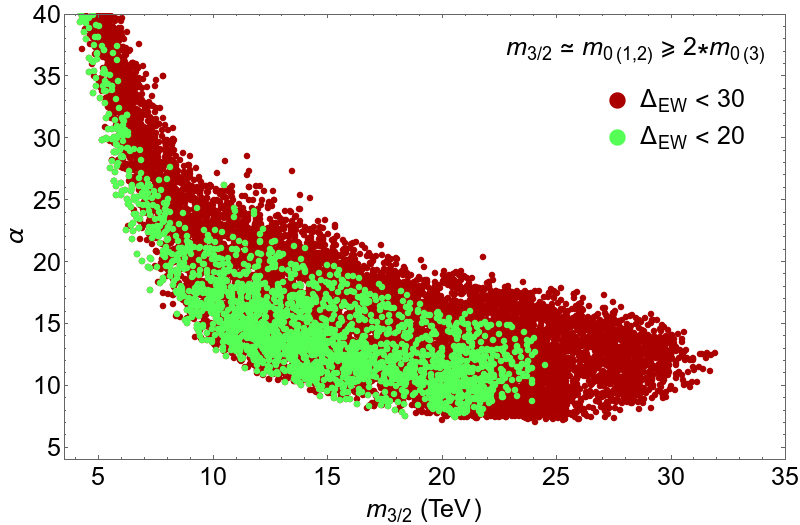}
\caption{Allowed SUSY  solutions in the $\alpha$ vs. $m_{3/2}$ plane 
of the natural mini-landscape  model with other parameters 
scanned over as described in the text. 
We take $c_m=(16\pi^2/\alpha )^2$ and $c_{m3} < c_m/4$ to enforce $m_0(1,2)\simeq m_{3/2}\gtrsim 2 m_0(3)$.
\label{fig:scan}
}
\end{center}
\end{figure}

A second result from Fig. \ref{fig:scan} is that we obtain a lower
bound on $\alpha \agt 7$. This bound arises from the LHC bound on
$m_{\tg}$ as can be seen from Fig.~\ref{fig:M3} It can be translated
via Eq.~(\ref{eq:mumir}) into a {\em lower bound} on the mirage
unification of $\mu_{\rm mir}\agt 2.7\times 10^{11}$ GeV. As a result, the
weak scale gaugino spectrum is somewhat compressed, but gross
compression is not possible.  This is relevant for
collider as well as for WIMP dark matter searches.

\subsubsection{The mini2 benchmark model} \label{subsubsec:mini2}

From Fig. \ref{fig:scan}, we now readily pick out natural
mini-landscape models with $m_{3/2}\sim 10-30$~TeV.  A particular
choice in shown in Table \ref{tab:bm} and labelled as {\it mini2}.  The
mini2 benchmark point has $m_i(1,2)\simeq m_{3/2}=20$ TeV while third
generation scalars lie at $m_i(3)\sim 5$ TeV.  The light stop mass is
suppressed both by renormalization effects from 
1. the large top Yukawa coupling and 2. large first/second generation 
scalar masses as well as 3. by large intragenerational mixing: 
thus, $m_{\tst_1}=1341.7$ GeV, nearly at  the maximal reach of 
HL-LHC\cite{ATLAS:2013hta}.  
The gluino mass is also close to the ultimate reach of HL-LHC.  And yet the model is
quite natural with $\delew=17.6$. Indeed, natural SUSY models
beyond the LHC reach are not difficult to find. The light higgsinos
though would be easily accessible to ILC.  The spectrum from the mini2
benchmark model  point is illustrated in Fig. \ref{fig:spect2}.
\begin{figure}[tbp]
\begin{center}
\includegraphics[height=0.4\textheight]{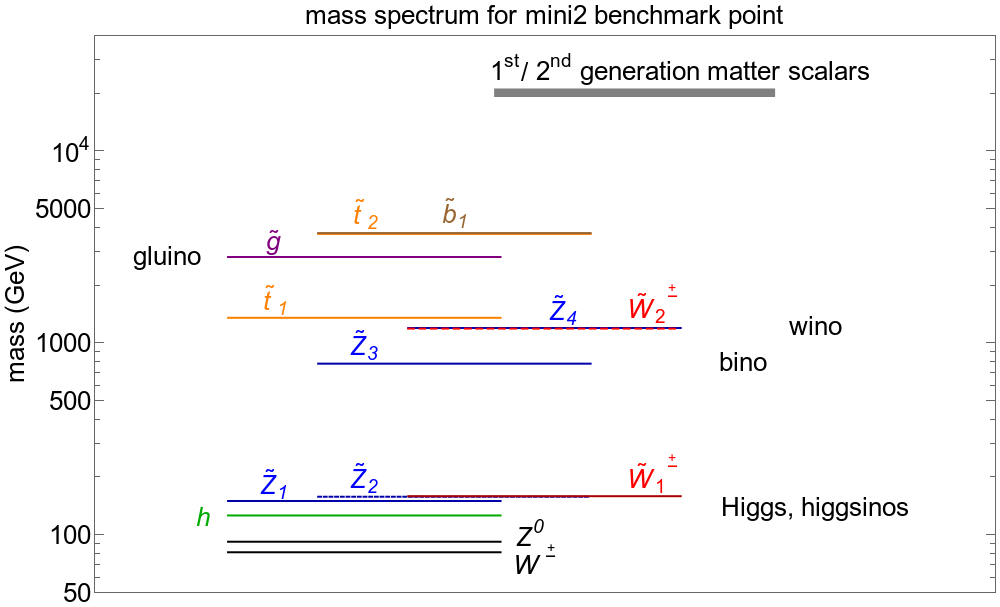}
\caption{The superparticle mass spectra from the
natural mini-landscape point mini2 of Table~\ref{tab:bm}.
\label{fig:spect2}}
\end{center}
\end{figure}

In Fig. \ref{fig:inos2} we show the evolution of gaugino masses from
the mini2 benchmark point. In this case, the mirage scale is clearly seen at
$\mu_{\rm mir}\sim 10^{13}$ GeV resulting in a mild compression of
gauginos as compared to models with gaugino mass unification.  Here,
we find $M_2/M_1\sim 1.5$ whereas -ino mass unification delivers
$M_2/M_1\sim 2$. Also, $M_3/M_1$ here is $\sim 3.6$ whereas unified
models tend to yield $M_3/M_1\sim 6$ (as in the NUHM3 BM point). In
obtaining these ratios, one must use the bino mass $m_{\tz_3}$ since
for natural SUSY the $\tw_1, \tz_{1,2}$ are all higgsino-like. Of
course, smaller values of $\alpha$ yield a greater compression of the
gaugino spectrum. We will return to this in Sec~\ref{subsec:extscan}
where we allow for deviations from Eq.~(\ref{eq:m1eqgrav}).
\begin{figure}[tbp]
\begin{center}
\includegraphics[height=0.39\textheight]{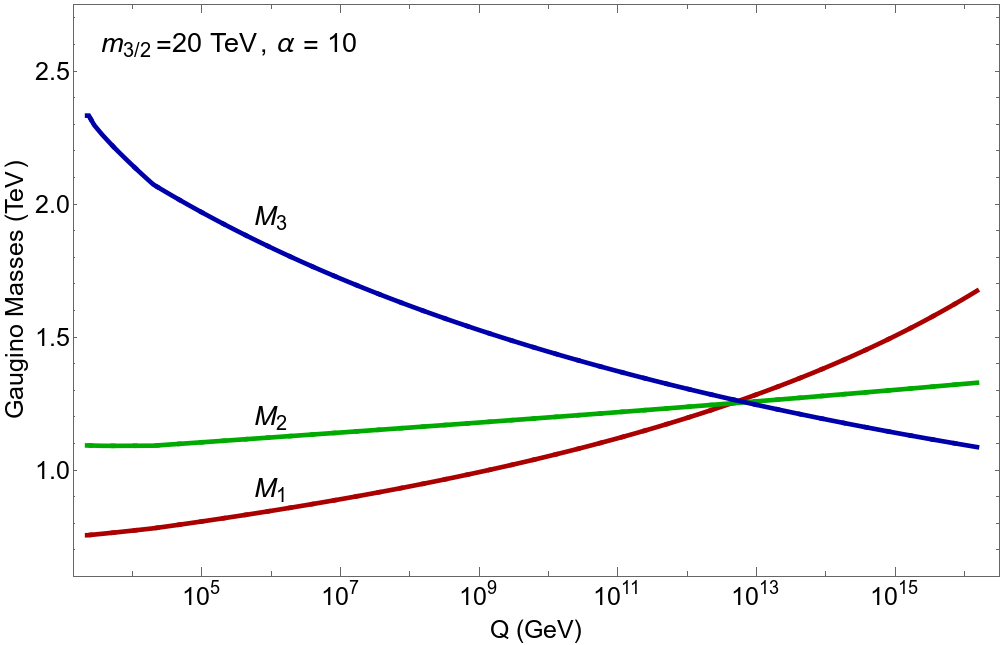}
\caption{Evolution of gaugino masses from the mini2 benchmark point with 
$m_{3/2}=20$ TeV and $\alpha =10$.
\label{fig:inos2}}
\end{center}
\end{figure}

In Fig. \ref{fig:scalars2}, we show the evolution of various soft
scalar masses for the mini2 benchmark point. The first/second
generation scalars lie at $\sim 20$ TeV and hardly run. Third
generation scalars lie around $5 $ TeV. The Higgs sector parameter
$m_{H_u}$ starts somewhat heavier than this at $Q=m_{\rm GUT}$ but is
radiatively-driven to natural low values at $Q\sim m_{weak}$ (notice
here that though $m_{H_u}^2$ does not run to a negative value, EWSB
nonetheless occurs once the negative radiative corrections $\Sigma_u^u$
are included).  The $\mu$ parameter hardly evolves and lies around
$\mu\sim 150$ GeV.  This figure illustrates well the three different
physical scales: $\mu\sim m_{weak}\sim 100$ GeV, $m(3,Higgs)\sim 5$
TeV and $m(1,2)\sim 20$ TeV.  We mention in passing that, in contrast
to the earliest MM models, the scalar evolution does not exhibit any
special feature at $Q=\mu_{\rm mir}$.
\begin{figure}[tbp]
\begin{center}
\includegraphics[height=0.39\textheight]{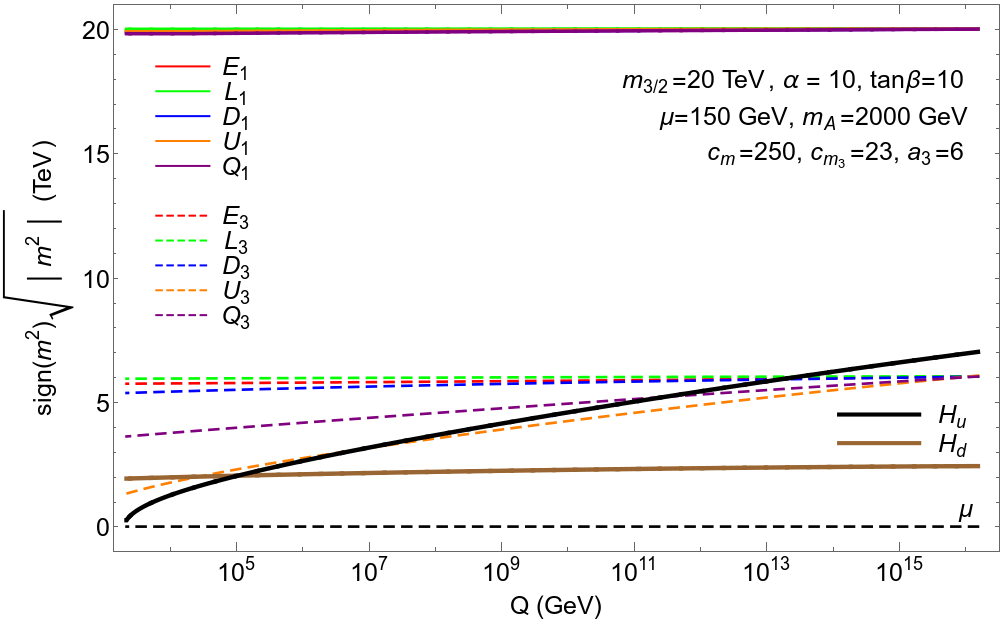}
\caption{Plot of running scalar masses from the mini2 benchmark point
with $m_{3/2}=20$ TeV, $\alpha =10$, $\tan\beta =10$ and $c_m=250$,
$a_3=6$ with $c_{m3}=23$, $\mu =150$ GeV and $m_A=2$ TeV at the weak scale.
\label{fig:scalars2}}
\end{center}
\end{figure}

\subsection{Effect of relaxing $m_0(1,2)\simeq m_{3/2}$} \label{subsec:extscan}

In Sec.~\ref{subsec:scan}, motivated by the
fact that the the first and second generation sfermion mass parameters
are less protected from SUSY breaking effects than the Higgs and top
squark multiplets, we had fixed $m_0(1,2)\simeq m_{3/2}$. This led us,
among other things, to conclude that the mirage scale could not be much
lower than $\sim 10^{11}$~GeV, with the associated mild compression of
the gaugino spectrum. Depending on the details of the location of the
first two generation fields, their SUSY breaking parameters may well be
partially protected so that $m_0(1,2)$ are somewhat smaller than
$m_{3/2}$ but, of course, still hierarchically separated from $m_0(3)$.

\begin{figure}[tbp]
\begin{center}
\includegraphics[height=0.4\textheight]{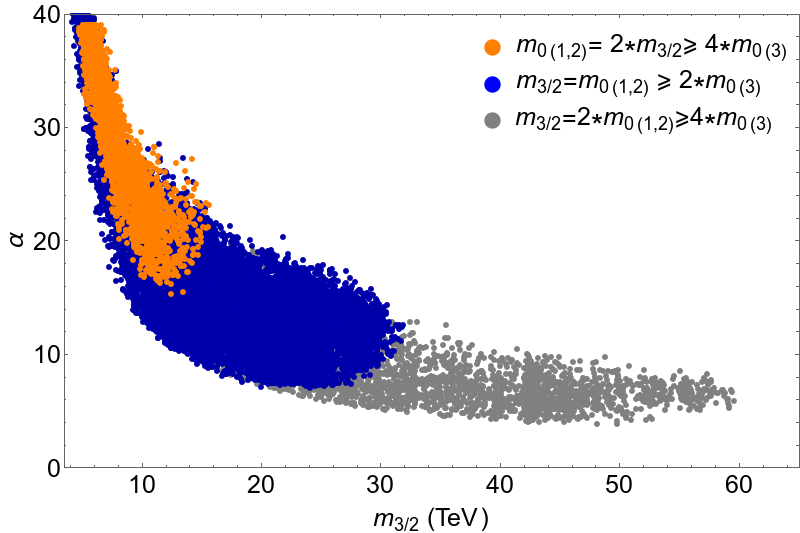}
\caption{Allowed SUSY solutions with $\delew <30$ 
  in the $\alpha$ vs. $m_{3/2}$ plane
  from an extended scan of the natural mini-landscape model with $c_m
  =4\times (16\pi^2/\alpha )^2$ to enforce $m_0(1,2)\simeq 2 m_{3/2}$ (orange
  points), $c_m =(16\pi^2/\alpha )^2$ to enforce $m_0(1,2)\simeq
  m_{3/2}$ (blue points) and $c_m =\frac{1}{4}(16\pi^2/\alpha )^2$
  to enforce $m_0(1,2)\simeq \frac{1}{2}m_{3/2}$ (gray points) as described
  in Sec.~\ref{subsec:extscan} of the text. 
  To maintain the hierarchy, we require 
  $m_0(3)<{\rm min}\left[ m_0(1,2)/2,\ m_{3/2}/2\right]$ in our scan. 
  Other parameters scanned over as in Fig.~\ref{fig:scan}. We note there are gray
  points not visible under the orange and blue dots extending down to
  low values of $m_{3/2}$. 
  \label{fig:genscan}
}
\end{center}
\end{figure}

Motivated by this, we adopt a phenomenological attitude and perform
other parameter scans, this time taking 
1. $m_0(1,2)\simeq m_{3/2}/2$ and 2. $m_0(1,2)\simeq 2m_{3/2}$. 
We also require that $m_0(1,2) \ge 2m_0(3)$ 
(and $m_{3/2} \ge 2m_0(3)$ in case \#2) to ensure that the hierarchy between
generations remains as a feature of the mini-landscape. 
The scanned range of other parameters is the same as in
Fig.~\ref{fig:scan}.  The solutions with $\delew$ from
this generalized scans that also satisfy the LHC constraints are
illustrated in Fig.~\ref{fig:genscan}. The blue dots show the same
results as in Fig.~\ref{fig:scan}. The gray dots show results for
the case where $m_0(1,2)\simeq \frac{1}{2}m_{3/2}$ while the orange dots
are for $m_0(1,2)\simeq 2m_{3/2}$. The main result is that for the
case with smaller values of $m_0(1,2)$ shown by the gray dots,
natural solutions with larger values of $m_{3/2}$ are allowed. This is
not surprising if we recall that the upper limit on $m_{3/2}$ comes
from the fact that the stop mass squared parameters become negative
due to two loop contributions involving correspondingly heavy squarks
in the first two generations. For a fixed gravitino mass, because 
$m_0(1,2)$ is about half as small for the gray points as compared with the
blue points, it is clear that there will be viable solutions out to
about twice larger gravitino masses in the gray point case. The
situation is exactly reversed for the $m_0(1,2)=2m_{3/2}$ case
illustrated by the orange points.

An important phenomenological consequence of the large $m_{3/2}$
solutions is that they extend to $\alpha$ values as small as 4, to be
compared with the bound $\alpha \agt 7$ that we saw from
Fig.~\ref{fig:scan}. As a result, the mirage unification scale can be
as low as $\sim 5\times 10^7$ GeV, with a concomitantly larger
compression of gauginos\footnote{ For instance, for a natural point
  with $m_{3/2}=50.6$~TeV and $\alpha= 4.3$ in the gray region, we
  have $M_1,M_2,M_3 = 1120, 1380, 2460$~GeV, respectively at the weak
  scale.}  relative to the situation in Fig.~\ref{fig:scan}.  While
our considerations emphasize that there is a {\em lower bound} on
$\mu_{\rm mir}$, the precise value of this lower bound is sensitively
dependent on just how small the ratio of $m_0(1,2)/m_{3/2}$ can be.

Before closing this discussion, we remind the reader that we were
motivated to do the extended scan in Fig.~\ref{fig:genscan} because
the first two generations may well not be located exactly at the
orbifold fixed point.  In this case they may have some partial
protection from SUSY breaking, resulting in soft terms smaller than
$m_{3/2}$, but not as small as those of the stop and Higgs
fields. From this perspective, the case with the orange dots is
disfavoured in the mini-landscape picture. We have nonetheless shown
it here for completeness.

\section{Implications for LHC, ILC  and dark matter searches}
\label{sec:col_dm}

In this section, we investigate briefly the prospects for discovery of
SUSY particles within the context of the natural mini-landscape
picture.  In this section, for brevity, all the results showing
$\delew$ versus the various sparticle masses are obtained for the
canonical case with $c_m=(16\pi^2/\alpha )^2$, so that $m_0(1,2)\simeq
m_{3/2}$, and requiring in addition that $m_0(1,2) \gtrsim 2m_0(3)$.
These plots have been made by merging the results of a broad scan
with those for a focussed scan for $\delew< 30$, and the range of
allowed values of $\mu$ is extended to 500~GeV.

\subsection{Consequences for LHC and LHC33}
\begin{figure}[tbp]
\begin{center}
\includegraphics[height=0.39\textheight]{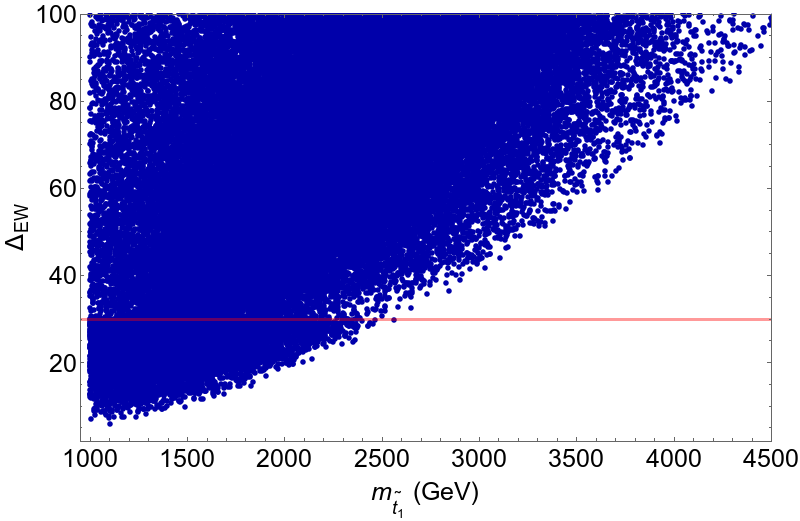}
\caption{Plot of $\delew$ vs. $m_{\tst_1}$ from a scan over 
natural mini-landscape parameter space with $m_0(1,2)\simeq m_{3/2}$.
\label{fig:del_mt1}}
\end{center}
\end{figure}

We begin by showing results for the value of $\delew$ vs. $m_{\tst_1}$
from our scan over mini-landscape parameter space in
Fig.~\ref{fig:del_mt1}. We see that for $\delew<20$ we expect
$m_{\tst_1}\alt 2$~TeV, while with the more conservative $\delew<30$
constraint $m_{\tst_1}$ may be as heavy as 2.5~TeV.  For SUSY mass
spectra from the natural mini-landscape where
$m_{\tw_1,\tz_{1,2}}\sim \mu\alt 200-300$ GeV, it has been found
that $B(\tst_1\to b\tw_1) \sim 50\%$ while $B(\tst_1\to t\tz_{1,2})$
are each at about 25\%\cite{Baer:2016bwh}. Meanwhile, the reach of
LHC14 for top-squark pair production in several simplified models has
been calculated in Ref. \cite{ATLAS:2013hta} and
\cite{Kim:2016rsd}. There, it was found that HL-LHC with $\sim 3$
ab$^{-1}$ of integrated luminosity, has a $5\sigma$ reach out to
$m_{\tst_1}\sim 1.1-1.4$ TeV. Apparently HL-LHC will cover only a
portion of mini-landscape parameter space via top squark pair
searches. Assuming that the lighter top squark decays via $\tst_1 \to
t\tz_{1,2}$ and $b\tw_1$ with branching ratios $\simeq 0.25, 0.25$ and
0.5, respectively, the entire allowed range of top squark masses in
Fig.~\ref{fig:del_mt1} should be accessible at LHC33 where the stop
reach extends to $m_{\tst_1} \sim 2.8$~TeV for an integrated
luminosity of 1~ab$^{-1}$ \cite{jamieprep}.

\begin{figure}[tbp]
\begin{center}
\includegraphics[height=0.39\textheight]{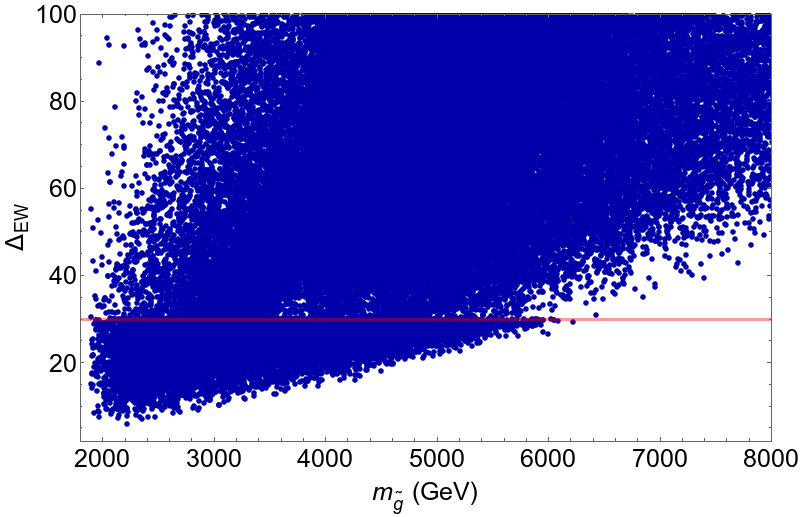}
\caption{Plot of $\delew$ vs. $m_{\tg}$ from a scan over 
natural mini-landscape parameter space with $m_0(1,2)\simeq m_{3/2}$.
\label{fig:del_mgl}}
\end{center}
\end{figure}
\begin{figure}[tbp]
\begin{center}
\includegraphics[height=0.39\textheight]{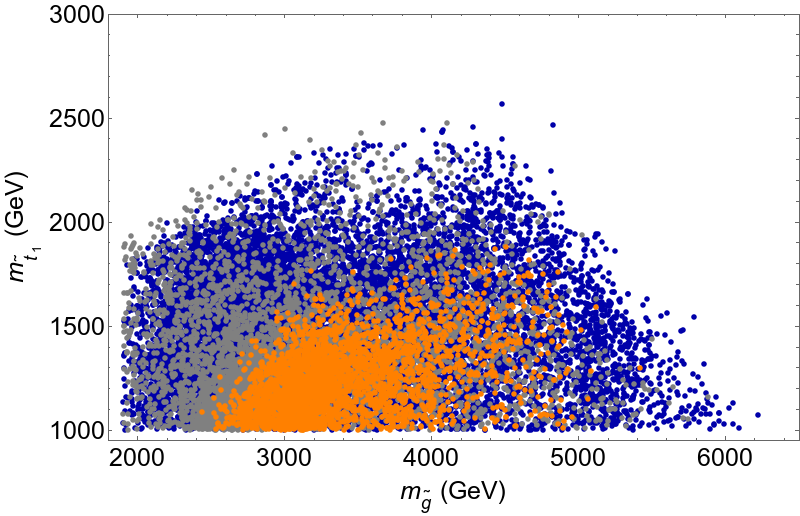}
\caption{A scatter plot of $m_{\tst_1}$ versus $m_{\tg}$ for solutions
  with $\delew < 30$ from the same three scans over the natural
  mini-landscape parameter space with $m_0(1,2) \simeq (2,1,1/2)\times
  m_{3/2}$ illustrated in Fig.~\ref{fig:genscan}. The color scheme in
  this figure is also the same as in Fig. \ref{fig:genscan}.
  \label{fig:extmgmstop} }
\end{center}
\end{figure}

In Fig. \ref{fig:del_mgl}, we plot the value of $\delew$ vs. $m_{\tg}$
from our scan over mini-landscape parameter space.  For $\delew<20$,
then $m_{\tg}\alt 4.5$ TeV while the more conservative bound
$\delew<30$ yields $m_{\tg}\alt 6$ TeV.  The upper bound on $m_{\tg}$
from the mini-landscape model is higher than the value
derived\cite{Baer:2015rja} from models such as NUHM2 where
$m_0(1,2)=m_0(3)$. This is because in the mini-landscape case the
positive contributions to stop masses from a heavy gluino (that lead to
the upper bound on its mass) are partially compenstated by the large
two-loop RGE contribution from heavy first/second generation scalars
which depress the stop mass parameters.
For the natural mini-landscape
spectra, usually $\tg\to t\tst_1$ followed by $\tst_1$ decays as
mentioned above.  The reach of HL-LHC has been calculated for
$pp\to\tg\tg X$ in Ref. \cite{Baer:2016wkz} where it was found that
the $5\sigma$ reach of LHC with 0.3 (3) ab$^{-1}$ extends to
$m_{\tg}\sim 2.4$ (2.8) TeV.  We see again that the HL-LHC will cover
only a portion of natural mini-landscape parameter space via gluino
pair searches. However, SUSY searches at
LHC33\cite{Baer:2017yqq,jamieprep} should be able to cover much of the
gluino range in Fig.~\ref{fig:del_mgl}.

In Fig.~\ref{fig:extmgmstop} we plot the points from the general scans 
in Fig.~\ref{fig:genscan} in the $m_{\tst_1}-m_{\tg}$ plane
using the same color coding as before. We see that the upper bounds on the stop
and gluino masses are insensitive to the precise value of
$m_0(1,2)/m_{3/2}$. 
Since the gluino reach of LHC33 extends to $\sim
5.5$~TeV if $m_{\tst_1}> 2$~TeV \cite{Baer:2017yqq}, we conclude that
LHC33 experiments should be sensitive to {\em both} gluino and squark
signals over most of the natural parameter space of the mini-landscape
framework, and of course, that SUSY will not evade detection at LHC33
if it is realized in this incarnation.

In Fig. \ref{fig:del_M2}, we plot $\delew$ versus the value of
the charged wino mass $m_{\tw_2}\simeq M_2({\rm weak})$. We see that for $\delew<20$
(30) then $m_{\tw_2}$ is bounded by $\sim 2$ (2.5) TeV.  This is
somewhat higher than the value obtained in models like NUHM2 with
gaugino mass unification where instead it is found that $m_{\tw_2}\alt
1.3$ (1.6) TeV\cite{Baer:2015rja}.  The wino mass is relevant to LHC
SUSY searches via the same-sign diboson channel where
$pp\to\tw_2\tz_4$ followed by $\tw_2\to W\tz_{1,2}$ and $\tz_4\to
W^\pm\tw_1^{\mp}$.
These decay modes lead about 50\% of the time to a
$W^\pm W^\pm +\eslt$ final state which provides a low jet activity
same-sign dilepton signature with very low SM backgrounds, the largest
of which  arises from $t\bar{t}W$ production. The LHC reach was
estimated in this channel for 1 ab$^{-1}$ to extend to $m_{1/2}\sim 1$
TeV corresponding to a reach in $m_{\tw_2}$ of about 0.85 TeV
\cite{Baer:2013yha}. A rough extrapolation to 3 ab$^{-1}$ should
extend HL-LHC reach to the vicinity of $m_{\tw_2}\sim 1.2$ TeV.  In
any case, again we see that HL-LHC can cover only a portion of natural
mini-landscape parameter space via the SSdB signature.
\begin{figure}[tbp]
\begin{center}
\includegraphics[height=0.39\textheight]{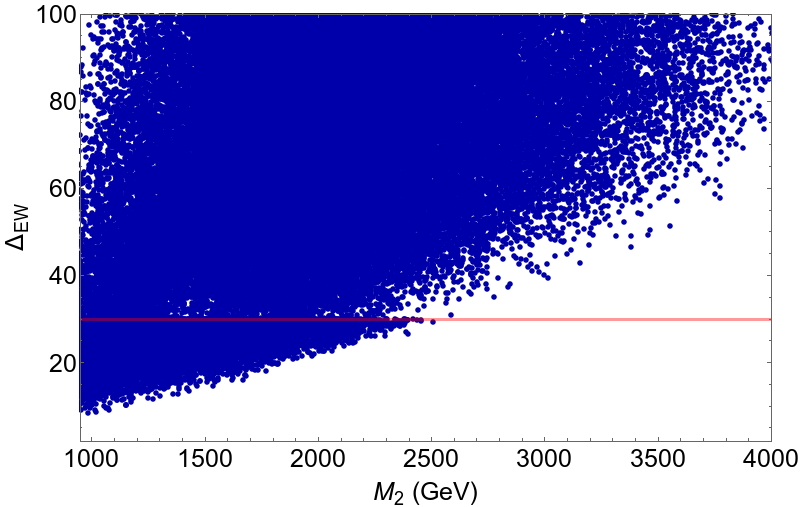}
\caption{Plot of $\delew$ vs. $M_2({\rm weak})$ from a scan over 
natural mini-landscape parameter space with $m_0(1,2)\simeq m_{3/2}$.
\label{fig:del_M2}}
\end{center}
\end{figure}

In the natural mini-landscape model, we expect the higgsinos to have
the tightest upper bounds from naturalness so that
$m_{\tw_1,\tz_{1,2}}\alt 200-300$ GeV. While higgsino pair production
can occur at large enough rates at LHC, the inter-higgsino mass gap is
small, {\it e.g.}  from Fig. \ref{fig:del_mz2mz1}, we see that
$m_{\tz_2}-m_{\tz_1}\sim 3-15$ GeV. As a result $\tz_2$, and
analogously also $\tw_1$, release very little visible energy in their
decays, and so mainly contribute to the missing transverse energy.
It has been shown that the resultant monojet signature from
$pp\to\tz_{1,2}\tz_{1,2}j$ or $\tw_1\tz_{1,2}j$ production at the LHC
(where $j$ denotes a hard QCD jet) occurs at only the 1-2\% level above
SM background from mainly $Zj$ production where $Z\to\nu\bar{\nu}$
\cite{mono}.  An alternative signature has been
suggested\cite{kribs,soft} where $pp\to\tz_1\tz_2j$ production followed
by $\tz_2\to\ell^+ \ell^- \tz_1$ giving rise to soft dileptons plus jet
(used for trigger) plus $\eslt$.  The reach of HL-LHC in this channel
has been found to extend to $\mu\sim 250$ GeV for mass gaps $\sim 10-20$
GeV.  In the case of the mini-landscape where bino and winos can be
somewhat heavier than in unified gaugino mass models the inter-higgsino
mass gap is typically smaller (less higgsino-gaugino mixing), as seen in
Fig. \ref{fig:del_mz2mz1}. This makes detection of the
$\ell^+\ell^-j+\eslt$ channel somewhat more difficult than in models
with gaugino mass unification both because the dilepton $p_T(\ell )$
spectra is softer and also because SM backgrounds from Drell-Yan and
$\Upsilon$ and associated production become more relevant.
\begin{figure}[tbp]
\begin{center}
\includegraphics[height=0.39\textheight]{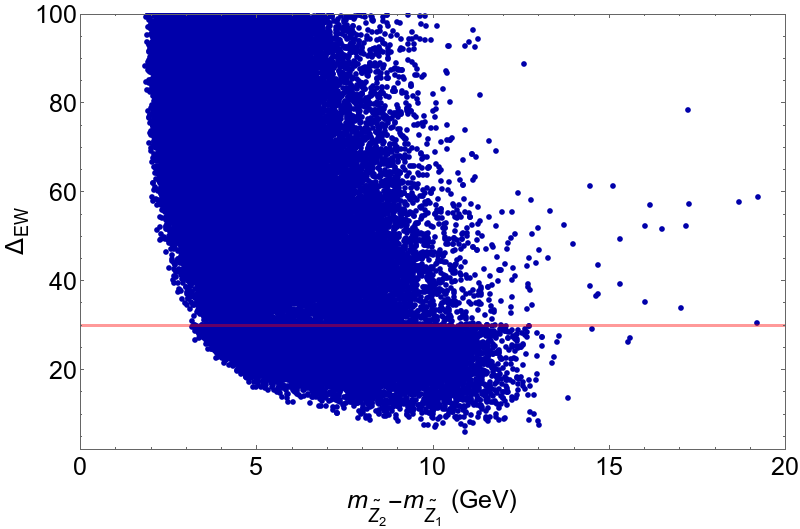}
\caption{Plot of $\delew$ vs. $m_{\tz_2}-m_{\ts_1}$ from a scan over 
natural mini-landscape parameter space with $m_0(1,2)\simeq m_{3/2}$.}
\label{fig:del_mz2mz1}
\end{center}
\end{figure}

\subsection{Consequences for ILC}

The proposed International Linear $e^+e^-$ Collider is proposed to be
built in Japan and could operate initially at $\sqrt{s}\sim 250$ GeV
as a Higgs factory with later upgrades to $\sqrt{s}=500$ and even 1000
GeV.  The light higgsinos $\tw_1$ and $\tz_{1,2}$ are required to be
not too far from $m_{W,Z,h}$ via the naturalness condition: see
Fig. \ref{fig:del_mw1} where for $\delew<20$ (30), we have
$m_{\tw_1}\alt 300$ (375) GeV.  This means that SUSY signals from
$e^+e^-\to\tw_1^+\tw_1^-$ and $\tz_1\tz_2$ processes should be
observable provided that these reactions are kinematically accessible.
The modest inter-higgsino mass gaps probably offer no great 
obstacle to discovery of higgsino pair production 
in the clean environment of $e^+e^-$
collisions\cite{ilc,Baer:2016new,Fujii:2017ekh}, although detailed
studies for mass gaps of $\sim 5-10$~GeV have not yet been completed.
\begin{figure}[tbp]
\begin{center}
\includegraphics[height=0.39\textheight]{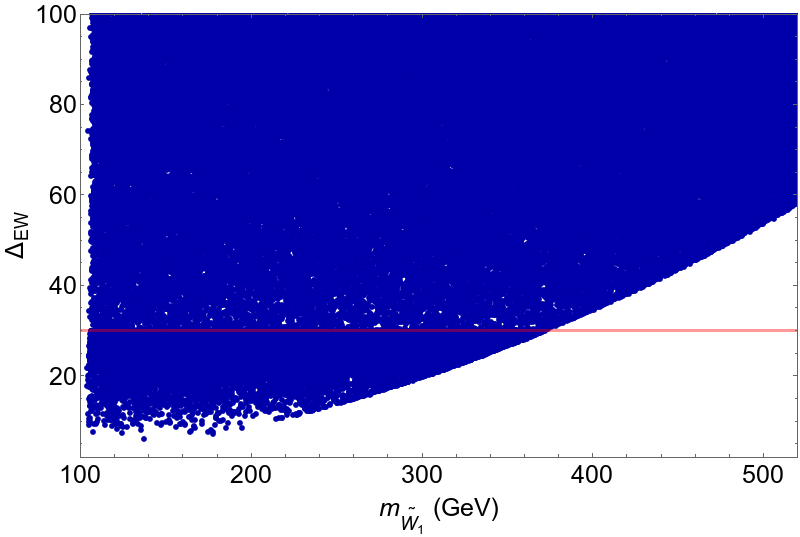}
\caption{Plot of $\delew$ vs. $m_{\tw_1}$ from a scan over 
natural mini-landscape parameter space with $m_0(1,2)\simeq m_{3/2}$.
\label{fig:del_mw1}}
\end{center}
\end{figure}

An additional benefit of $e^+e^-\to \tw_1^+\tw_1^-$ and $\tz_1\tz_2$
production is the precision measurements of $m_{\tw_1}$, $m_{\tz_2}$
and $m_{\tz_1}$. These measurements shoud give high precision on the
value of the superpotential $\mu$ parameter. Also, the inter-higgsino
mass splitting is dependent on the values of the gaugino masses $M_1$
and $M_2$.  From Ref's. \cite{Baer:2016new} and \cite{Fujii:2017ekh},
these ought to be extractable using fitting procedures. It would be
interesting to carefully examine whether these methods that have been
shown to provide useful measurements in a case study with a neutralino
mass gap of 22~GeV continue to work for the smaller mass gaps of
$3-15$~GeV typical of the mini-landscape picture.

Once the gaugino masses are extracted (including $M_3$ if gluino
pair production is found at LHC\cite{Baer:2016wkz} or its energy
upgrade) then one will be able to test if the gaugino masses 
unify at $Q=m_{\rm GUT}$, or at $Q=\mu_{\rm mir}$ as expected in the mini-landscape
picture where the gaugino mass pattern is as given by 
mirage mediation.

\subsection{Consequences for WIMP and axion searches}

Dark matter in the natural mini-landscape framework is expected to occur
as a mixture of QCD axions and higgsino-like WIMPs.  The WIMPs are
thermally underproduced owing to large higgsino-higgsino annihilation
and co-annihilation reactions in the early universe.  Typically the
higgsino-like WIMP thermal abundance is a factor 10-20 below the
measured value. Since it is reasonable to require naturalness in the QCD
sector as well (solving the strong CP problem), 
the QCD axion is a highly motivated candidate for the remaining dark matter. 
The SUSY DFSZ axion has been suggested as a
solution to the SUSY $\mu$ problem\cite{Kim:1983dt} while simultaneously
allowing for a little hierarchy\cite{radpq} $\mu\sim f_a^2/m_{\rm Pl}\ll
m_{SUSY}\sim \Lambda^3/m_{\rm Pl}^2$ where $\Lambda$ is the scale for
gaugino condensation occuring in the hidden sector.

While axions are produced as usual non-thermally via vacuum
mis-alignment, one must also account for the other components of the
axion superfield: the spin-$1/2$ axino $\ta$ and the spin-0 saxion
$s$. Axinos and saxions are expected to acquire masses $\sim
m_{3/2}\sim 10-50$ TeV. Axinos can be produced thermally and if they
decay after WIMP freeze-out then they augment the WIMP
abundance. Saxions can be produced thermally but also via coherent
oscillations.  If they decay after freeze-out, then they also may
augment the WIMP abundance.  If they decay dominantly to SM particles
then they may inject late time entropy into the cosmic plasma thus
diluting any relics which are present.  And if they decay to axions
$s\to aa$ then they may inject additional relativistic degrees of
freedom in the cosmic plasma (for which there are strong bounds on
additional neutrino species $\Delta N_{\rm eff}\alt 1$).  The exact
abundances of axions and higgsino-like WIMPs depends on the various PQ
sector parameters and sample calculations are shown in the eight
coupled Boltzmann equation solutions from Ref. \cite{Bae:2014rfa}. It
was found that for much of the allowed parameter space, axions
dominate the relic abundance\cite{Bae:2013bva}.

Prospects for higgsino-like WIMP direct detection via spin-independent
(SI) or spin-dependent (SD) scattering have been shown in
Ref. \cite{Baer:2016ucr} for a variety of models.  A key point here is
that the detection rates may be lowered by up to a factor $\xi\equiv
\Omega_{\tz_1}h^2/0.12$ to account for the depleted local abundance of
WIMPs from the usually assumed density $\rho_{\rm local}\simeq
0.3$~GeV/cm$^3$. The indirect WIMP detection rates from cosmic WIMP
annihilation depend on the square of the WIMP density, and so are
suppressed by a factor of $\xi^2$.

\begin{figure}[tbp]
\begin{center}
\includegraphics[height=0.4\textheight]{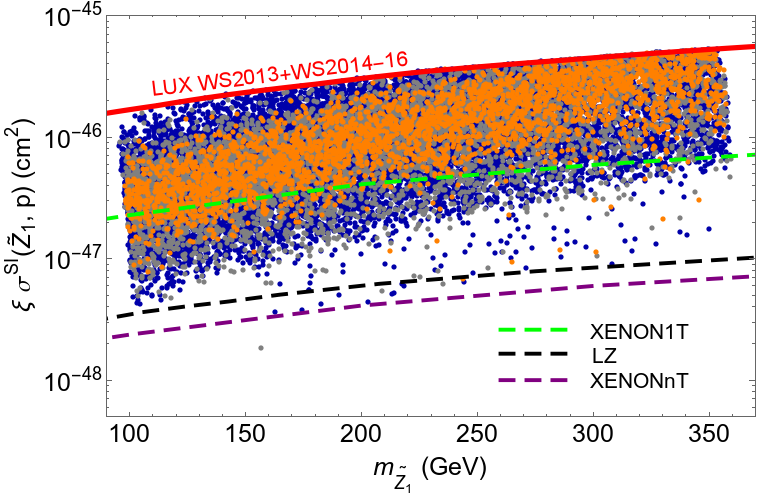}
\caption{A scatter plot of $\xi\sigma^{SI}(\tz_1,p)$ versus $m_{\tz1}$
  for solutions with $\delew < 30$ from the same three scans over the
  natural mini-landscape parameter space with $m_0(1,2) \simeq
  (2,1,1/2)\times m_{3/2}$ illustrated in Fig.~\ref{fig:genscan}. Here
  $\xi$ is the higgsino fraction of the total CDM density, assuming that
  the relic density of higgsino WIMPs is given by its thermal value. The
  colour scheme in this figure is the same as in
  Fig. \ref{fig:genscan}.
  \label{fig:SIsigma} }
\end{center}
\end{figure}
In Fig.~\ref{fig:SIsigma}, we show a plot of the expected scaled
spin-independent WIMP-nucleon cross section for natural mini-landscape
models assuming that the relic density of the higgsino-like WIMP is
given by its thermal value.  We show results for the same three scans from
Fig.~\ref{fig:genscan}, using the same colour-coding as in this
figure. We plot only those points consistent with the current bounds
from the LUX experiment (with 95+332 live days combined exposure) \cite{lux}. 
The expected direct detection rates are not very sensitive to the
ratio $m_0(1,2)/m_{3/2}$, assuming it is within a factor 2 of unity.  We
also show projections for the sensitivity of the XENON1T, XENONnT
\cite{xenon} and the LZ \cite{lz} experiments. In contrast to
expectations in natural SUSY models with gaugino mass unification where
it was concluded that XENON1T would be sensitive to the direct detection
signal over the bulk of parameter space \cite{Baer:2016ucr}, we see that
for the mini-landscape picture multi-ton detectors will be needed for
complete coverage. This is because the bino and wino masses can
be substantially larger in the mini-landscape picture compared to models
with unified gaugino masses, reducing the gaugino content in the
higgsino-like $\tz_1$. As a result, the $h\tz_1\tz_1$ coupling which
arises only via gaugino-higgsino-Higgs boson interactions, is
correspondingly reduced. Since WIMP-nucleon scattering is typically
dominated by the $h$-exchange contribution, the direct detection cross
section can decrease to smaller values in the mini-landscape picture. It
is heartening though that future detectors such as LZ and XENONnT 
are projected to probe the entire natural mini-landscape 
parameter space subject to
the usual caveats that there is no injection of entropy (from late
decays of moduli or from the decays of saxions) that dilutes the WIMP
density below its thermal value.

Turning to indirect detection, we have also evaluated expectations for
detection of gamma ray signals from cosmic WIMP annihilation. 
We find that these are a factor
of 10-20 below the current bounds from the Fermi-LAT/MAGIC 
collaboration\cite{fermi}, assuming WIMP annihilation to $W^+W^-$
pairs. We also find that, except perhaps at the highest values of WIMP
masses in the last figure, the gamma ray signals also lie beyond the
reach of the CTA \cite{cta}, assuming a 500 h exposure. We do not show
these results for the sake of brevity.

\section{Conclusions}
\label{sec:conclude}

In this paper, we have examined the superparticle mass spectra and 
broad phenomenological features of the mini-landscape picture, 
focussing on the region of parameter
space consistent with electroweak naturalness. The mini-landscape
scenario is a string-motivated construction based on the expectation
that the MSSM emerges as the low energy theory in special regions of
the string landscape. The salient feature of this scenario is that the
multiplet structure as well as the masses of the MSSM fields depends
on their location in the compactified manifold. 
The symmetry group of the low energy theory is just $G_{\rm SM}$, but first and second
generation multiplets that live near the orbifold fixed point have
enhanced symmetry and enter as complete representations of $SO(10)$,
while only the SM gauge, Higgs and third generation matter multiplets
remain at lower energies.  Superymmetry breaking is also felt
differently by the various particles.  Gaugino mass parameters are
suppressed relative to $m_{3/2}$ and exhibit the mirage mediation
pattern in Eq.~(\ref{eq:Ma}). Third generation squark and soft Higgs
parameters are also relatively suppressed, while first/second
generation soft SUSY breaking masses are expected to be comparable to
$m_{3/2}$. The mini-landscape picture leads to the parametrization of
soft SUSY breaking parameters given by Eqs.~(\ref{eq:Ma}-\ref{eq:MHd})
which we have dubbed generalized mirage mediation. This framework is
completely specified by the parameter set (\ref{eq:gmmp}).

We have identified the region of model parameter space consistent with
low electroweak fine-tuning $\delew \le 30$. 
The $\delew$ measure yields the most conservative value of fine tuning 
in the sense that it allows for the fact that soft SUSY breaking parameters -- 
that are often regarded as independent -- 
may actually be correlated by the SUSY breaking mechanism. 
The main features of the superparticle spectra in this
preferred region are summarized below and compared to corresponding
features of the natural NUHM2 model where gaugino mass unification is assumed.

\bi
\item As in all models with low electroweak fine-tuning, we expect
  light higgsino states $\tz_{1,2},\tw_1^{\pm}$ with masses not far
  above $m_Z$. In the mini-landscape scenario, the neutral higgsino
  mass splitting is typically 3-15~GeV (to be compared with 10-25~GeV
  in the natural NUHM2 model\cite{Baer:2015rja}).

\item In contrast to the NUHM2 model where gaugino mass unification
  leads to weak scale gaugino masses in the ratio $M_1:M_2:M_3 \simeq
  1:2:6$, the gaugino spectrum from the natural mini-landscape 
  may be compressed. The degree of
  compression sensitively depends on the mirage unification scale
  $\mu_{\rm mir}$ which in turn depends on how low $\alpha$ can be
  while maintaining consistency with LHC bounds on $m_{\tg}$. 
  This compression is
  relatively mild if we assume the squark mass parameters of the first
  two generations are close to $m_{3/2}$ but significantly larger
  compression is possible if these squarks are much lighter
  than $m_{3/2}$.

\item We find that $m_{\tg} \alt 6$~TeV and $m_{\tst_1} \alt 2.5$~TeV
  if $\delew < 30$ and $m_0(1,2)$ is within a factor 2 of
$m_{3/2}$. Moreover, $m_{\tg}> 5$~TeV only when $m_{\tst_1}< 2$~TeV.

\item The first two generations of squarks and sleptons are very
  heavy. While this puts them well beyond the range of LHC, it
  also ameliorates the SUSY flavour problem.

  \ei

While it is possible that experiments at the LHC may discover the gluino
or the top squark if SUSY is realized in the natural region of
mini-landscape parameters, their discovery is not guaranteed at even the
HL-LHC. Moreover, the discovery of SUSY via $W^\pm W^\pm + \eslt$ events
which is nearly guaranteed at the HL-LHC in the natural NUHM2 model, is no
longer a sure thing within the mini-landscape picture because the
compression of the gaugino spectrum now allows much heavier winos 
even in natural models. 
This same compression also leads to a reduced mass
difference $m_{\tz_2}-m_{\tz_1}$ rendering the mono-jet plus soft
dilepton signal (which was observable in the natural NUHM2 model) more
difficult to extract. We are thus forced to conclude that SUSY
detection is not guaranteed over the entire natural parameter space of
the mini-landscape model even at the HL-LHC. Detection
of higgsino-like WIMPs at XENON1T is also not guaranteed in the
mini-landscape picture. Larger detectors such as XENONnT and LZ will,
however cover the entire natural mini-landscape parameter space unless
late injection of entropy reduces the WIMP density from what we expect
assuming that the higgsino is a thermally produced relic in standard Big
Bang cosmology.


Turning to future colliders, it is very likely that experiments at
electron positron colliders 
will be able to detect higgsinos via $e^+e^- \to \tz_1\tz_2,
\tw_1^+\tw_1^-$ production if these reactions are kinematically
accessible.  Experiments at LHC33 will be able to access top squark
signals over the entire natural SUSY mass range, and also gluino signals
over almost all of the natural range of $m_{\tg}$ in the mini-landscape
scenario.

\section*{Acknowledgments}
We thank Peter Nilles for several communications and clarifications about the
mini-landscape scenario.  We  thank Werner Porod for checking our
GMM boundary conditions using Spheno.  This work was supported in
part by the US Department of Energy, Office of High Energy Physics.
The computing for this project was performed at the OU Supercomputing
Center for Education \& Research (OSCER) at the University of Oklahoma
(OU).

%

%

\begin{thebibliography}{99}
\small
%

\bibitem{Schellekens:2013bpa}
  A.~N.~Schellekens,
  Rev.\ Mod.\ Phys.\  {\bf 85} (2013) no.4,  1491
  [arXiv:1306.5083 [hep-ph]];
K.~R.~Dienes,
  Phys.\ Rept.\  {\bf 287} (1997) 447
  [hep-th/9602045].

\bibitem{Dienes:2015xua}
  K.~R.~Dienes,
  Adv.\ Ser.\ Direct.\ High Energy Phys.\  {\bf 22} (2015) 81.

\bibitem{Nilles:2015wua}
  H.~P.~Nilles and P.~K.~S.~Vaudrevange,
  Adv.\ Ser.\ Direct.\ High Energy Phys.\  {\bf 22} (2015) 49.

\bibitem{Witten:1981nf}
  E.~Witten,
  Nucl.\ Phys.\ B {\bf 188} (1981) 513;
E.~Witten,
  Phys.\ Lett.\  {\bf 105B} (1981) 267;
R.~K.~Kaul,
  Phys.\ Lett.\  {\bf 109B} (1982) 19.

\bibitem{gauge} S.~Dimopoulos, S.~Raby and F.~Wilczek,
  Phys.\ Rev.\ D {\bf 24} (1981) 1681;
U.~Amaldi, W.~de Boer and H.~Furstenau,
  Phys.\ Lett.\ B {\bf 260}, 447 (1991);
J.~R.~Ellis, S.~Kelley and D.~V.~Nanopoulos,
  Phys.\ Lett.\ B {\bf 260} (1991) 131;
P.~Langacker and M.~x.~Luo,
  Phys.\ Rev.\ D {\bf 44} (1991) 817.
%
\bibitem{rewsb} L. E. Iba\~nez and G. G. Ross, Phys. Lett. {\bf B110}, 215
(1982); K. Inoue {\it et al.} Prog. Theor. Phys. {\bf 68}, 927 (1982)
and {\bf 71}, 413 (1984);
L.~Iba\~nez, Phys. Lett. {\bf B118}, 73 (1982);
 H.~P.~Nilles, M.~Srednicki and D.~Wyler,
  Phys.\ Lett.\ B {\bf 120} (1983) 346;
J.~Ellis, J.~Hagelin, D.~Nanopoulos and M.~Tamvakis,
Phys. Lett. {\bf B125}, 275 (1983);
L.~Alvarez-Gaum\'e. J.~Polchinski and M.~Wise,
Nucl. Phys. {\bf B221}, 495 (1983);
B.~A.~Ovrut and S.~Raby,
  Phys.\ Lett.\ B {\bf 130} (1983) 277;
for a review, see
L.~E.~Ibanez and G.~G.~Ross,
  Comptes Rendus Physique {\bf 8} (2007) 1013.
%
\bibitem{lhc_h} G.~Aad {\it et al.} [ATLAS Collaboration],
  Phys.\ Lett.\ B {\bf 716} (2012) 1;
S.~Chatrchyan {\it et al.} [CMS Collaboration],
  Phys.\ Lett.\ B {\bf 716} (2012) 30.
%
\bibitem{mhiggs} H.~E.~Haber and R.~Hempfling,
  Phys.\ Rev.\ Lett.\  {\bf 66} (1991) 1815;
J.~R.~Ellis, G.~Ridolfi and F.~Zwirner,
  Phys.\ Lett.\ B {\bf 257} (1991) 83;
Y.~Okada, M.~Yamaguchi and T.~Yanagida,
  Prog.\ Theor.\ Phys.\  {\bf 85} (1991) 1;
For a review, see {\it e.g.}  M.~S.~Carena and H.~E.~Haber,
  Prog.\ Part.\ Nucl.\ Phys.\  {\bf 50} (2003) 63
  [hep-ph/0208209].
%
\bibitem{Cleaver:1998im}
  G.~Cleaver, M.~Cvetic, J.~R.~Espinosa, L.~L.~Everett and P.~Langacker,
  Nucl.\ Phys.\ B {\bf 545} (1999) 47.
%
\bibitem{Cvetic:2002qa}
  M.~Cvetic, P.~Langacker and G.~Shiu,
  Phys.\ Rev.\ D {\bf 66} (2002) 066004.
%
\bibitem{Kane:2006yi}
  G.~L.~Kane, P.~Kumar and J.~Shao,
  J.\ Phys.\ G {\bf 34} (2007) 1993.
%
\bibitem{Lebedev:2006kn}
  O.~Lebedev, H.~P.~Nilles, S.~Raby, S.~Ramos-Sanchez, M.~Ratz, P.~K.~S.~Vaudrevange and A.~Wingerter,
  Phys.\ Lett.\ B {\bf 645} (2007) 88
  [hep-th/0611095];
O.~Lebedev, H.~P.~Nilles, S.~Ramos-Sanchez, M.~Ratz and P.~K.~S.~Vaudrevange,
  Phys.\ Lett.\ B {\bf 668} (2008) 331
  [arXiv:0807.4384 [hep-th]].

\bibitem{inocond} H.~P.~Nilles,
  Phys.\ Lett.\  {\bf 115B} (1982) 193.
H.~P.~Nilles,
  Nucl.\ Phys.\ B {\bf 217} (1983) 366.
S.~Ferrara, L.~Girardello and H.~P.~Nilles,
  Phys.\ Lett.\  {\bf 125B} (1983) 457.
for a review, see 
H.~P.~Nilles,
  hep-th/0402022.

\bibitem{Kobayashi:2004ya}
  T.~Kobayashi, S.~Raby and R.~J.~Zhang,
  Nucl.\ Phys.\ B {\bf 704} (2005) 3.
%

\bibitem{Nilles:2015qka}
  H.~P.~Nilles,
  Adv.\ High Energy Phys.\  {\bf 2015} (2015) 412487.

  %
  \bibitem{Krippendorf:2012ir}
  S.~Krippendorf, H.~P.~Nilles, M.~Ratz and M.~W.~Winkler,
  Phys.\ Lett.\ B {\bf 712} (2012) 87
  [arXiv:1201.4857 [hep-ph]];
  M.~Badziak, S.~Krippendorf, H.~P.~Nilles and M.~W.~Winkler, J.~High Energy Phys. {\bf 1303} 094 (2013).
  %
\bibitem{amsb} L. Randall and R. Sundrum, Nucl. Phys. {\bf B557}, 79 (1999);
G. F. Giudice, M. Luty, H. Murayama and R. Rattazzi, 
J. High Energy Phys.{\bf 9812}, 027 (1998);
J. Bagger, T. Moroi and E. Poppitz, 
J. High Energy Phys. {\bf 0004}, 009 (2000);
P. Binetruy, M. K. Gaillard and B. Nelson, Nucl. Phys. {\bf B604}, 32 (2001).
%
\bibitem{choi} K. Choi, A. Falkowski, H. P. Nilles, M. Olechowski and
S. Pokorski, J. High Energy Phy{\bf 0411}, 076 (2004); 
K. Choi, A. Falkowski, H. P. Nilles
and M. Olechowski, Nucl. Phys. {\bf B718}, 113 (2005).
J.~P.~Conlon, F.~Quevedo and K.~Suruliz, JHEP {\bf 0508}, 007 (2005)
  [arXiv:hep-th/0505076];
A.~Pierce and J.~Thaler,
  JHEP {\bf 0609} (2006) 017;
B.~L.~Kaufman, B.~D.~Nelson and M.~K.~Gaillard,
  Phys.\ Rev.\ D {\bf 88} (2013) no.2,  025003.
%
\bibitem{kklt} S. Kachru, R. Kallosh, A. Linde and S. P. Trivedi,
Phys. Rev. {\bf D68}, 046005 (2003).
%
\bibitem{uplift} O.~Lebedev, H.~P.~Nilles and M.~Ratz,
  Phys.\ Lett.\ B {\bf 636} (2006) 126
  [hep-th/0603047];
K.~A.~Intriligator, N.~Seiberg and D.~Shih,
  JHEP {\bf 0604} (2006) 021
  [hep-th/0602239];
M.~Gomez-Reino and C.~A.~Scrucca,
  JHEP {\bf 0605} (2006) 015
  [hep-th/0602246];
E.~Dudas, C.~Papineau and S.~Pokorski,
  JHEP {\bf 0702} (2007) 028
  [hep-th/0610297];
H.~Abe, T.~Higaki, T.~Kobayashi and Y.~Omura,
  Phys.\ Rev.\ D {\bf 75} (2007) 025019
  [hep-th/0611024];
O.~Lebedev, V.~Lowen, Y.~Mambrini, H.~P.~Nilles and M.~Ratz,
  JHEP {\bf 0702} (2007) 063
  [hep-ph/0612035];
H.~Abe, T.~Higaki, T.~Kobayashi and Y.~Omura,
  JHEP {\bf 0804} (2008) 072
  [arXiv:0801.0998 [hep-th]];
V.~Lowen, H.~P.~Nilles and A.~Zanzi,
  Phys.\ Rev.\ D {\bf 78} (2008) 046002
  [arXiv:0804.3913 [hep-th]].
%
\bibitem{Lowen:2008fm}
  V.~Lowen and H.~P.~Nilles,
  Phys.\ Rev.\ D {\bf 77} (2008) 106007
  [arXiv:0802.1137 [hep-ph]].
%
\bibitem{choi3} K. Choi, K-S. Jeong and K. Okumura, 
J. High Energy Phys. {\bf 0509}, 039 (2005).
%
\bibitem{flm} A. Falkowski, O. Lebedev and Y. Mambrini, 
J. High Energy Phys. {\bf 0511}, 034 (2005).
%
\bibitem{cjko} K.~Choi, K.~S.~Jeong, T.~Kobayashi and K.~i.~Okumura,
  Phys.\ Rev.\ D {\bf 75} (2007) 095012.
%
\bibitem{Baer:2012up} H.~Baer, V.~Barger, P.~Huang, A.~Mustafayev and X.~Tata,
  Phys.\ Rev.\ Lett.\  {\bf 109} (2012) 161802
  [arXiv:1207.3343 [hep-ph]].
%
\bibitem{Baer:2012cf} H.~Baer, V.~Barger, P.~Huang, D.~Mickelson, A.~Mustafayev and X.~Tata,
  Phys.\ Rev.\ D {\bf 87} (2013) no.11,  115028
  [arXiv:1212.2655 [hep-ph]].
%
\bibitem{Baer:2014ica}
  H.~Baer, V.~Barger, D.~Mickelson and M.~Padeffke-Kirkland,
  Phys.\ Rev.\ D {\bf 89} (2014) no.11,  115019
  [arXiv:1404.2277 [hep-ph]].
%
\bibitem{Baer:2016hfa}
  H.~Baer, V.~Barger, H.~Serce and X.~Tata,
  Phys.\ Rev.\ D {\bf 94} (2016) no.11,  115017
  [arXiv:1610.06205 [hep-ph]].
  %
\bibitem{Barbieri:1987fn} R.~Barbieri and G.~F.~Giudice,
  Nucl.\ Phys.\ B {\bf 306} (1988) 63.
%
\bibitem{Kitano:2006gv} R.~Kitano and Y.~Nomura,
  Phys.\ Rev.\ D {\bf 73} (2006) 095004
  [hep-ph/0602096].
%
\bibitem{Papucci:2011wy} M.~Papucci, J.~T.~Ruderman and A.~Weiler,
  JHEP {\bf 1209} (2012) 035
  [arXiv:1110.6926 [hep-ph]].
%
\bibitem{Baer:2013gva}
  H.~Baer, V.~Barger and D.~Mickelson,
  Phys.\ Rev.\ D {\bf 88} (2013) no.9,  095013
  [arXiv:1309.2984 [hep-ph]].
%
\bibitem{Mustafayev:2014lqa}
  A.~Mustafayev and X.~Tata,
  Indian J.\ Phys.\  {\bf 88} (2014) 991
  [arXiv:1404.1386 [hep-ph]].
%
\bibitem{Chan:1997bi}
  K.~L.~Chan, U.~Chattopadhyay and P.~Nath,
  Phys.\ Rev.\ D {\bf 58} (1998) 096004;
%
\bibitem{Baer:2011ec} H.~Baer, V.~Barger and P.~Huang,
  JHEP {\bf 1111} (2011) 031
  [arXiv:1107.5581 [hep-ph]].

%
\bibitem{choi_sax} K.~Choi and K.~S.~Jeong,
  JHEP {\bf 0701} (2007) 103.
%
\bibitem{Douglas:2012bu}
  M.~R.~Douglas,
  arXiv:1204.6626 [hep-th].
%
\bibitem{Kim:1983dt}
  J.~E.~Kim and H.~P.~Nilles,
  Phys.\ Lett.\  {\bf 138B} (1984) 150.
%
  %
\bibitem{radpq} 
K.~J.~Bae, H.~Baer and H.~Serce,
  Phys.\ Rev.\ D {\bf 91} (2015) no.1,  015003.
%
\bibitem{landscape} H.~Baer, V.~Barger, M.~Savoy and H.~Serce,
  Phys.\ Lett.\ B {\bf 758} (2016) 113.
%
\bibitem{Choi:2007ka}
  K.~Choi and H.~P.~Nilles,
  JHEP {\bf 0704} (2007) 006
  [hep-ph/0702146 [HEP-PH]].
%
\bibitem{nuhm2}  D.~Matalliotakis and H.~P.~Nilles,
  Nucl.\ Phys.\ B {\bf 435} (1995) 115;
P.~Nath and R.~L.~Arnowitt,
  Phys.\ Rev.\ D {\bf 56} (1997) 2820;
J. Ellis, K. Olive and Y. Santoso, \plb{539}{2002}{107};
J. Ellis, T. Falk, K. Olive and Y. Santoso, \npb{652}{2003}{259};
H.~Baer, A.~Mustafayev, S.~Profumo, A.~Belyaev and X. Tata, \jhep {0507}{2005}{065}.
%
\bibitem{isajet}  ISAJET 7.86, by H.~Baer, F.~Paige, S.~Protopopescu and
X.~Tata, \hepph{0312045}; 
Isasugra, by H.~Baer, C.~H.~Chen, R.~B.~Munroe, F.~E.~Paige and X.~Tata,
  Phys.\ Rev.\ D {\bf 51} (1995) 1046.
%
\bibitem{atlas_s} The ATLAS collaboration [ATLAS Collaboration],
  ATLAS-CONF-2017-022.
%
\bibitem{cms_s} A.~M.~Sirunyan {\it et al.} [CMS Collaboration],
  arXiv:1704.07781 [hep-ex].
%
\bibitem{Baer:2017yqq}
  H.~Baer, V.~Barger, J.~S.~Gainer, P.~Huang, M.~Savoy, H.~Serce and X.~Tata,
  arXiv:1702.06588 [hep-ph].
%
\bibitem{h125} H.~Baer, V.~Barger and A.~Mustafayev,
  Phys.\ Rev.\ D {\bf 85} (2012) 075010.
  %
\bibitem{Bae:2013bva}
  K.~J.~Bae, H.~Baer and E.~J.~Chun,
  Phys.\ Rev.\ D {\bf 89} (2014) no.3,  031701
  [arXiv:1309.0519 [hep-ph]];
%
\bibitem{mv}   S.~P.~Martin and M.~T.~Vaughn,
  Phys.\ Rev.\ D {\bf 50} (1994) 2282
   Erratum: [Phys.\ Rev.\ D {\bf 78} (2008) 039903]
  [hep-ph/9311340].
%
\bibitem{Baer:2000xa}
  H.~Baer, C.~Balazs, P.~Mercadante, X.~Tata and Y.~Wang,
  Phys.\ Rev.\ D {\bf 63} (2001) 015011
  doi:10.1103/PhysRevD.63.015011
  [hep-ph/0008061].


\bibitem{Giudice:2006sn}
  G.~F.~Giudice and R.~Rattazzi,
  Nucl.\ Phys.\ B {\bf 757} (2006) 19
  [hep-ph/0606105].
%
\bibitem{atlas_t1} The ATLAS collaboration [ATLAS Collaboration],
  ATLAS-CONF-2017-020.
%
\bibitem{cms_t1} CMS Collaboration [CMS Collaboration],
  CMS-PAS-SUS-16-051.
%
\bibitem{Dine:1990jd}
  M.~Dine, A.~Kagan and S.~Samuel,
  Phys.\ Lett.\ B {\bf 243} (1990) 250.
%
\bibitem{Khlopov:1984pf}
  M.~Y.~Khlopov and A.~D.~Linde,
  Phys.\ Lett.\  {\bf 138B} (1984) 265.
  doi:10.1016/0370-2693(84)91656-3
%
\bibitem{Kohri:2005wn}
  K.~Kohri, T.~Moroi and A.~Yotsuyanagi,
  Phys.\ Rev.\ D {\bf 73} (2006) 123511
  [hep-ph/0507245].
%
\bibitem{Baer:2016bwh}
  H.~Baer, V.~Barger, N.~Nagata and M.~Savoy,
  Phys.\ Rev.\ D {\bf 95} (2017) no.5,  055012,
   arXiv:1611.08511 [hep-ph].
%
\bibitem{ATLAS:2013hta}
  ATLAS Collaboration,
  ATLAS-PHYS-PUB-2013-011.
%
\bibitem{Kim:2016rsd}
  J.~S.~Kim, K.~Rolbiecki, R.~Ruiz, J.~Tattersall and T.~Weber,
  arXiv:1606.06738 [hep-ph].
  %
  \bibitem{jamieprep} H.~Baer, V.~Barger, J.~Gainer, M.~Savoy,
    D.~Sengupta and X.~Tata, paper in preparation.
  %
\bibitem{Baer:2016wkz}
  H.~Baer, V.~Barger, J.~S.~Gainer, P.~Huang, M.~Savoy, D.~Sengupta and X.~Tata,
  arXiv:1612.00795 [hep-ph].
%
\bibitem{Baer:2015rja}
  H.~Baer, V.~Barger and M.~Savoy,
  Phys.\ Rev.\ D {\bf 93} (2016) no.3,  035016
  [arXiv:1509.02929 [hep-ph]].
%
\bibitem{Baer:2013yha}
  H.~Baer, V.~Barger, P.~Huang, D.~Mickelson, A.~Mustafayev, W.~Sreethawong and X.~Tata,
  Phys.\ Rev.\ Lett.\  {\bf 110} (2013) no.15,  151801
  [arXiv:1302.5816 [hep-ph]];
H.~Baer, V.~Barger, P.~Huang, D.~Mickelson, A.~Mustafayev, W.~Sreethawong and X.~Tata,
  JHEP {\bf 1312} (2013) 013
   Erratum: [JHEP {\bf 1506} (2015) 053]
  [arXiv:1310.4858 [hep-ph]].
  %
\bibitem{mono} C.~Han, A.~Kobakhidze, N.~Liu, A.~Saavedra, L.~Wu and
  J.~M.~Yang,
  \jhep{1402}{2014}{049}
  [arXiv:1310.4274 [hep-ph]]; 
  H.~Baer, A.~Mustafayev and X.~Tata,
  \prd{89}{2014}{055007}
  [arXiv:1401.1162 [hep-ph]];
  P.~Schwaller and J.~Zurita, 
  \jhep{1403}{2014}{060} [arXiv:1312.7350 [hep-ph]]; 
  %
\bibitem{kribs} Z.~Han, G.~D.~Kribs, A.~Martin and A.~Menon,
  \prd{89}{2014}{075007}
      [arXiv:1401.1235~[hep-ph]].
%
\bibitem{soft} H.~Baer, A.~Mustafayev and X.~Tata, \prd{90}{2014}{115007}.
  %
\bibitem{ilc} H.~Baer, V.~Barger, D.~Mickelson, A.~Mustafayev and X.~Tata,
  JHEP {\bf 1406} (2014) 172.
%
\bibitem{Baer:2016new}
  H.~Baer, M.~Berggren, K.~Fujii, S.~L.~Lehtinen, J.~List, T.~Tanabe and J.~Yan,
  arXiv:1611.02846 [hep-ph].
%
\bibitem{Fujii:2017ekh}
  K.~Fujii {\it et al.},
  arXiv:1702.05333 [hep-ph].

\bibitem{Bae:2014rfa}
  K.~J.~Bae, H.~Baer, A.~Lessa and H.~Serce,
  JCAP {\bf 1410} (2014) no.10,  082
  [arXiv:1406.4138 [hep-ph]].

K.~J.~Bae, H.~Baer and E.~J.~Chun,
  JCAP {\bf 1312} (2013) 028
  [arXiv:1309.5365 [hep-ph]].
%
\bibitem{Baer:2016ucr}
  H.~Baer, V.~Barger and H.~Serce,
  Phys.\ Rev.\ D {\bf 94} (2016) no.11,  115019
  [arXiv:1609.06735 [hep-ph]].
%
\bibitem{lux} D.~Akerib {\it et al.} (LUX collaboration)
 \prl{118}{2017}{021303} [arXiv:1608.07648 [astro-ph.CO]].
%
\bibitem{xenon} E.~Aprile {\it et al.} JCAP {\bf 1604} (2016) 027
  [arViv: 1512.07501 [phys.ins-det]].
%
\bibitem{lz} D.~Akerib {\it et al.}, LZ Conceptual Design Report,
arXiv:1509.02910 [physics.ins-det].
%
\bibitem{fermi} M.~Ahnen {\it et al.} (Fermi-LAT/MAGIC
  Collaboration) JCAP {\bf 1602} (2016) 039
  [arXiv:1601.06590[astro.ph-HE]].
%
\bibitem{cta}M.~Wood, J.~Buckley, S.~Digel, S.~Funk, D.~Nieto and
  M.~Sanchez-Conde, {\em Proc. Community Summer Study 2013: Snowmass on
  the Mississippi (CSS2013): Minneapolis, MN, July-Aug. (2013)},
  [arXiv:1305.0302[astro-ph.HE]].


\end{thebibliography}
\end{document}